\documentclass{book}
\usepackage{qpt_book}
\usepackage{amsfonts,amsmath,amssymb} 
\usepackage{graphicx}
\usepackage{epstopdf}
\usepackage{wasysym}

\newcommand{\bra}[1]{\mbox{$\langle #1 |$}}
\newcommand{\ket}[1]{\mbox{$| #1 \rangle$}}

\newcommand{\proj}[1]{\mbox{$|#1\rangle \!\langle #1 |$}}
\newcommand{\GS}[0]{\Psi_{\mbox{\tiny GS}}}
\newcommand{\gs}[0]{\mbox{\tiny GS}}
\newcommand{\tr}[0]{\mbox{tr}}

\begin{document}

\pagestyle{myheadings}

%%%%%%%%%%%%%%%%%%%%%%%%%%%%%%%%%%%%%%%%%%%%%%%%%%%%%%%%%%%%%%%%%%%%%%%%%%%%%%%

\markboth{Understanding Quantum Phase Transitions}
{Entanglement Renormalization: an introduction}

\chapter{Entanglement Renormalization:\\ an introduction}

\vskip -0.5cm
{\bf \large Guifre Vidal} \\
{\it School of Mathematics and Physics, the University of Queensland, \\ Brisbane, QLD 4072, Australia} \\

\noindent
We present an elementary introduction to entanglement renormalization, a real space renormalization group for quantum lattice systems. This manuscript corresponds to a chapter of the book "Understanding Quantum Phase Transitions", edited by Lincoln D. Carr (Taylor $\&$ Francis, Boca Raton, 2010)
%%%%%%%%%%%%%%%%%%%%%%%%%%%%%%%%%%%%%%%%%%%%%%%%%%%%%%%%

\section{Introduction}

One of the main goals of Physics is to identify and characterize the possible phases of matter, as well as the transitions between these phases. In this chapter we are concerned with theoretical and computational aspects of quantum phases and quantum phase transitions, involving extended quantum many-body systems at zero temperature. 

Specifically, we consider the problem of constructing a real-space \emph{coarse-graining transformation} for quantum lattice systems at zero temperature. One such transformation should dispose of the degrees of freedom corresponding to small distances while preserving the low energy properties of the system. Obtaining an effective description of a system in terms of less degrees of freedom is obviously very convenient for numerical calculations. Our priority here is, however, to explore the use of the coarse-graining transformation within the context of the \emph{renormalization group} (RG) \cite{V:Kadanoff67, V:Wilson75, V:Fisher98, V:Cardy02}. Accordingly, we aim to define a RG flow in the space of local Hamiltonians, and to study quantum phases and quantum phase transitions by characterizing the fixed points of this flow.

The coarse-graining scheme that we will describe is known as \emph{entanglement renormalization}. Entanglement renormalization has been shown to be suitable to address the emergence of different types of order in systems of quantum spins in one and two spatial dimensions, including symmetry-breaking order and topological order, as well as to characterize quantum critical points. In addition, the formalism can be generalized to study models where the basic degrees of freedom are fermionic or, more generally, anyonic. We will introduce the approach step by step throughout the chapter, and will then apply it to the characterization of quantum critical points.

The content is organized in sections as follows. Sect. \ref{V:sect:rsRG} discusses a coarse-graining transformation, based on \emph{isometries}, that is capable of preserving ground state properties, but that fails to remove some of the short-distance degrees of freedom, which accumulate over successive iterations. This degrees of freedom are associated to short-range entanglement in the ground state of the system. Failure to properly get rid of short-range entanglement has important numerical implications, and precludes the use of RG ideas within this first coarse-graining scheme.

Sect. \ref{V:sect:ER} explains how to remove short-range entanglement from the lattice. This is achieved by introducing \emph{disentanglers} that act across the boundary of blocks of sites before the coarse-graining step. In this way \emph{all} short-distance degrees of freedom are removed from the system. Local operators are seen to be mapped into local operators by the \emph{ascending superoperator} $\mathcal{A}$, while its dual, the \emph{descending superoperator} $\mathcal{D}$, allows us to recover a fine-grained reduced density matrix from a coarse-grained one. We also describe the class of states that can be exactly represented using the entanglement renormalization scheme, known as the \emph{multi-scale entanglement renormalization ansatz} (MERA).

Sect. \ref{V:sect:RG} connects the present approach with the RG formalism. A RG map $\mathcal{R}(h)$ in the space of two-site interactions $h$ is defined in terms of the ascending superoperator $\mathcal{A}$. Each fixed point of the resulting (discrete) RG flow corresponds to a ground state that is invariant under the coarse-graining transformation and is described by a \emph{scale invariant} MERA. The entanglement of scale invariant ground states allows to distinguish between different types of fixed points. At the fixed point, the ascending superoperator, now independent of the length scale, is known as the scaling superoperator $\mathcal{S}$. 

Sect. \ref{V:sect:QuCrit} applies entanglement renormalization to the study of quantum phase transitions. Many universal properties of a quantum critical point, including critical exponents and the conformal data characterizing the pertinent conformal field theory, can be extracted from the scaling superoperator $\mathcal{S}$. We also address boundary critical phenomena by adding a boundary to the scale invariant MERA.

The formalism of the entanglement renormalization and the MERA were introduced in Refs. \cite{V:Vidal07,V:Vidal08}. 
Algorithms to approximate ground states have been described in Refs. \cite{V:Dawson08,V:Rizzi08,V:Evenbly09}. 
Two dimensional systems have been explored in Refs. 
\cite{V:Evenbly07, V:Evenbly08, V:Cincio08, V:Evenbly09b, V:Evenbly09c,V:Aguado08,V:Koenig08},  
including scalable simulations in interacting systems \cite{V:Evenbly09b, V:Evenbly09c} and analytical results for systems with topological order \cite{V:Aguado08,V:Koenig08}. 
Scale invariant systems, including non-critical and critical fixed points of the RG flow, have been studied in Refs. 
\cite{V:Vidal07, V:Vidal08, V:Evenbly09, V:Evenbly07, V:Evenbly08, V:Aguado08, V:Koenig08, V:Giovannetti08, V:Pfeifer09, V:Montangero08, V:Giovannetti09, V:Evenbly09d}.  
Extensions to two-dimensional systems with fermionic and anyonic degrees of freedom have been proposed in Refs. \cite{V:Corboz09, V:Corboz09b, V:Eisert09, V:Eisert09b, V:Aguado09}. 

%%%%%%%%%%%%%%%%%%%%%%%%%%%%%%%%%%%%%%%%%%%%%%%%%%%%%%%%%%%%%%%%%%%%%%%%%%%%%%%%%%%%%%%%%%%%%
%%%%%%%%%%%%%%%%%%%%%%%%%%%%%%%%%%%%%%%%%%%%%%%%%%%%%%%%%%%%%%%%%%%%%%%%%%%%%%%%%%%%%%%%%%%%%

\section{Coarse-graining and ground state entanglement}
\label{V:sect:rsRG}

Let us consider a system, such as a quantum spin model, that can be described by a lattice $\mathcal{L}$ in $D$ spatial dimensions. For simplicity, most of the present derivations will involve a lattice in one spatial dimension. However, one of the highlights of entanglement renormalization is that it also applies to higher dimensional cases, to which we will refer occasionally throughout the chapter. 

The microscopic degrees of freedom are placed on the $N$ sites of $\mathcal{L}$, with each site being described by a vector space $\mathbb{V}$ of finite dimension $d$. The model is further characterized by a Hamiltonian $H$ that decomposes as the sum of local terms, that is, of terms that act on a small number of neighboring sites. Except for the discussion on boundary critical phenomena in Sect. \ref{V:sect:QuCrit}, we assume that the model is invariant under translations. 

Our ultimate goal is to be able to compute low energy properties of the system, which we will assume to be in the ground state $\ket{\GS} \in \mathbb{V}^{\otimes N}$ of $H$. Let $o_1, o_2, \cdots, o_k$ be arbitrary local operators acting on different parts of the lattice. Then we would like to compute quantities such as 
\begin{equation}
	\langle o_1 o_2 \cdots o_k \rangle_{\GS} \equiv \bra{\GS} o_1 o_2 \cdots o_k \ket{\GS},
\label{V:eq:oGS}
\end{equation}
since from these expected values one can predict how the system reacts to arbitrary external probes\footnote{We will only consider equal-time correlators.}. However, due to the exponential growth in $N$ of the dimension of $\mathbb{V}^{\otimes N}$, computing $\ket{\GS}$ by diagonalizing $H$ is only affordable for very small systems. In order to address larger systems, we need a better plan. The strategy that we will pursue here is to build a transformation that removes short-distance degrees of freedom from the lattice model $(\mathcal{L},H)$, which is mapped into an effective lattice model $(\mathcal{L}',H')$ such that
\begin{equation}
	\langle o'_1 o'_2 \cdots o'_k \rangle_{\GS'} = \langle o_1 o_2 \cdots o_k \rangle_{\GS},
\label{V:eq:oGS'}
\end{equation}
where $\ket{\GS'}$ is the ground state of $H'$ and the operators $o'_1,o'_2, \cdots, o'_k$ result from transforming $o_1,o_2,\cdots, o_k$. 

Building an effective description can be helpful in several ways. On the one hand, in the case of a finite system, where the model $(\mathcal{L}',H')$ has a smaller Hilbert space dimension than the original model, diagonalizing $H'$ is computationally more affordable than diagonalizing $H$. As a result, larger models might be addressed. In conjunction with finite size scaling techniques, this already constitutes a powerful numerical route to study e.g. quantum critical phenomena. On the other hand, and closer to the goals of this chapter, a coarse-graining transformation that properly removes short-distance degrees of freedom could be used to investigate how $H$ changes under scale transformations. Then, with the help of powerful RG ideas, we might be able to evaluate Eq. \ref{V:eq:oGS} directly in the thermodynamic limit.

\subsection{A real space coarse-graining transformation}

%%%%%%%%%%%%%%%%%%%%%%%%%%%%%%%%
\begin{figure}
\begin{center}
\includegraphics[width=10cm]{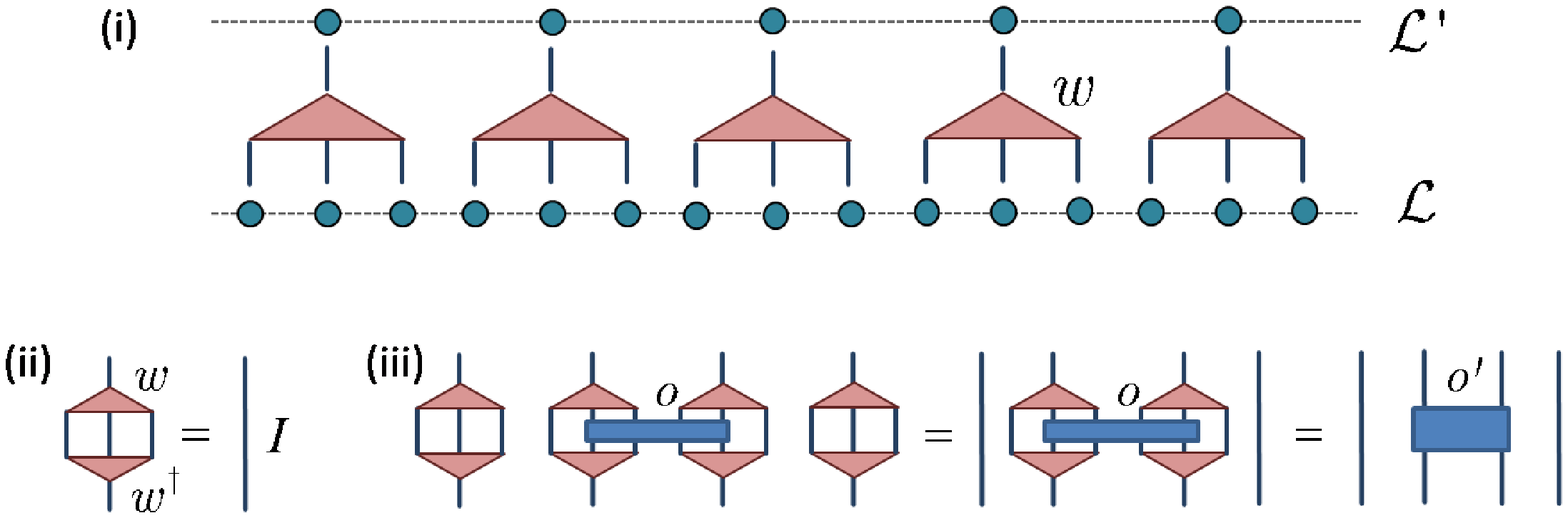}
\end{center}
\caption{(i) Coarse-graining transformations characterized by an isometry $w$ that maps blocks of three sites of lattice $\mathcal{L}$ into single sites of a coarse-grained lattice $\mathcal{L}'$. (ii) Graphical representation of $w^{\dagger} w=I_{\mathbb{V}'}$. (iii) An operator $o$ supported on e.g. two blocks of sites of $\mathcal{L}$ becomes a two-site operator $o'$, cf. Eq. \ref{V:eq:oo}. 
} 
\label{V:fig:isometry}
\end{figure}
%%%%%%%%%%%%%%%%%%%%%%%%%%%%%%%%%

Following the seminal works of Migdal, Kadanoff and Wilson in real space RG \cite{V:Kadanoff67, V:Wilson75, V:Fisher98}, we will proceed by coarse-graining blocks of sites of $\mathcal{L}$ into single sites of $\mathcal{L'}$. For concreteness, we divide $\mathcal{L}$ in blocks of three sites and, as in Wilson's numerical renormalization group (NRG) approach, we implement the coarse-graining by means of an isometry $w$,
\begin{equation}
	w: \mathbb{V}' \longmapsto \mathbb{V}^{\otimes 3},~~~~~~w^{\dagger}w = I_{\mathbb{V}'}, ~~ww^{\dagger} \equiv P,~~ P^2 = P,
	\label{V:eq:w}
\end{equation}
where $\mathbb{V}^{\otimes 3}$ is the vector space of the three sites, $\mathbb{V}'$ is the vector space of the effective site, $I_{\mathbb{V}'}$ is the identity operator in $\mathbb{V}'$ and $P$ is a projector onto the subspace of $\mathbb{V}^{\otimes 3}$ that is preserved by the coarse-graining. Fig. \ref{V:fig:isometry} illustrates this transformation, which defines an effective lattice $\mathcal{L}'$ made of $N'=N/3$ sites, as well as an effective Hamiltonian $H'$ given by
\begin{equation}
	H' = W^{\dagger} H W,~~~~~~~~~~~W \equiv w^{\otimes N/3}.
\end{equation}
A local operator $o$ with support on $r$ blocks of $\mathcal{L}$ becomes a local operator $o'$ supported on $r$ sites of $\mathcal{L}'$ given by
\begin{equation}
	o \rightarrow o' = \left(w^{\dagger\otimes r}\right) o \left(w^{\otimes r}\right),
	\label{V:eq:oo}
\end{equation}
where the isometries act on the relevant sites of $\mathcal{L}$. Notice that the support of local operators may shrink under coarse-graining, but it never expands. In particular, if $H$ can be expressed as a sum of interactions between pairs of nearest neighbor sites, then $H'$ will also contain at most nearest neighbor interactions.

The above transformation may not preserve the properties of the original ground state, in the sense of Eq. \ref{V:eq:oGS'}. As explained by White as part of his density matrix renormalization group (DMRG) \cite{V:White92}, in order to preserve ground state properties, the isometry $w$ must retain the whole support of the ground state reduced density matrix $\rho$ on the three-site block,
\begin{equation}
\rho \equiv \mbox{tr}_{E} \proj{\GS},
\end{equation}
where $E$ denotes all the sites of $\mathcal{L}$ not contained in the block. Let 
\begin{equation}
	\rho = \sum_{\alpha=1}^{\chi} p_{\alpha} \proj{\Psi_{\alpha}},~~~~~ \sum_{\alpha} p_{\alpha} = 1, ~~~p_{1} \geq p_{2} \geq \cdots > 0,
\label{V:eq:rho}
\end{equation}
be the eigenvalue decomposition of $\rho$. Then \emph{White's rule} consists in choosing $w$ such that the projector $P$ in Eq. \ref{V:eq:w} corresponds to the support of $\rho$, 
\begin{equation}
	P =  ww^{\dagger} = \sum_{\alpha=1}^{\chi} \proj{\Psi_{\alpha}}.~~~~~~~~~\mbox{White's rule}
\label{V:eq:proj}
\end{equation}
In particular, the dimension of $\mathbb{V}'$ is the number $\chi$ of non-zero eigenvalues\footnote{In practical calculations, one often neglects contributions from the eigenvectors $\ket{\Psi_{\alpha}}$ with smallest weights $p_{\alpha}$, at the price of introducing small errors in Eq. \ref{V:eq:oGS'}. Most of the subsequent discussions also apply in the case of approximate coarse-graining transformations.} of $\rho$. If Eq. \ref{V:eq:proj} holds, then it can be seen that the ground state $\ket{\GS'}$ of $H'$ is given by $\ket{\GS'} = W^{\dagger}\ket{\GS}$, whereas $WW^{\dagger} \ket{\GS} = \ket{\GS}$. It follows that\footnote{For simplicity, we assume that the local operators $o'_1, o'_2, \cdots, o'_k$ are supported on different sites of $\mathcal{L}'$, so that $W^{\dagger} o_1 o_2 \cdots o_k W = o'_1 o'_2 \cdots o'_3$. If e.g. operators $o'_1$ and $o'_2$ had overlapping support, then they would be fused into a single local operator $o' = \left(w^{\dagger\otimes r}\right) o_1 o_2 \left(w^{\otimes r}\right)$ supported on some larger number $r$ of sites.} 
\begin{eqnarray}
		\bra{\GS} o_1 o_2 \cdots o_k \ket{\GS} 
		= \bra{\GS} (WW^{\dagger}) o_1 o_2 \cdots o_k (WW^{\dagger})\ket{\GS} \\
		= \left(\bra{\GS} W\right) \left( W^{\dagger} o_1 o_2 \cdots o_k W\right) \left(W^{\dagger}\ket{\GS}\right)
		= \bra{\GS'} o'_1 o'_2 \cdots o'_k \ket{\GS'},
\label{V:eq:proof}
\end{eqnarray}
which indeed amounts to Eq. \ref{V:eq:oGS'}.

By iteration, from the original model $(\mathcal{L}^{(0)},H^{(0)}) \equiv (\mathcal{L},H)$ we can now build a sequence of increasingly coarse-grained lattice models 
\begin{eqnarray}
	(\mathcal{L}^{(0)}, H^{(0)}) \stackrel{w^{(1)}}{\longrightarrow} (\mathcal{L}^{(1)}, H^{(1)}) \stackrel{w^{(2)}}{\longrightarrow} (\mathcal{L}^{(2)}, H^{(2)}) \rightarrow \cdots,
\end{eqnarray}
where lattice $\mathcal{L}^{(\tau+1)}$ results from coarse-graining lattice $\mathcal{L}^{(\tau)}$, such that ground state properties are preserved at each step,
\begin{equation}
	\langle o^{(\tau+1)}_1 o^{(\tau+1)}_2 \cdots o^{(\tau+1)}_k \rangle_{\GS^{(\tau+1)}} = \langle o^{(\tau)}_1 o^{(\tau)}_2 \cdots o^{(\tau)}_k \rangle_{\GS^{(\tau)}}.
\label{V:eq:oGSRG}
\end{equation}

How well does this coarse-graining scheme remove short-distance degrees of freedom? Is it viable in practice? To answer these questions, it is useful to consider the sequence of ground state reduced density matrices $\{\varrho^{(0)}, \varrho^{(1)}, \varrho^{(2)}, \cdots\}$ corresponding to increasingly large blocks of sites of $\mathcal{L}$, where $\varrho^{(\tau)}$ corresponds to $3^{\tau}$ sites. White's rule, which guarantees that ground state properties are preserved, also implies that the vector space dimension $\chi^{(\tau)}$ of a site of lattice $\mathcal{L}^{(\tau)}$ must be given by the rank of $\varrho^{(\tau)}$. Thus, our next step is to characterize the sequence of effective dimensions $\{\chi^{(0)}, \chi^{(1)}, \chi^{(2)}, \cdots\}$.
 
\subsection{Ground state entanglement}

The rank $\chi$ of the ground state reduced density matrix $\rho$ of a block of sites depends on the amount of entanglement between that block of sites and the rest of the lattice, as can be seen from the so-called Schmidt decomposition, which expands the ground state $\ket{\GS}$ in terms of the eigenvectors $\ket{\Psi_{\alpha}}$ of $\rho$ (Eq. \ref{V:eq:rho}) and some orthonormal set of states $\ket{\Phi_{\alpha}}$ for the rest of the lattice,
\begin{equation}
	\ket{\GS} = \sum_{\alpha=1}^{\chi} \sqrt{\lambda_{\alpha}} \ket{\Psi_{\alpha}}\otimes \ket{\Phi_{\alpha}}.
\end{equation}

When $\chi=1$, corresponding to a reduced density matrix $\rho = \proj{\Psi_1}$ in a pure state, the ground state $\ket{\GS} = \ket{\Psi_1}\otimes \ket{\Phi_1}$ factorizes into the product of individual states for the block and for the rest of the lattice. Instead, if $\chi>1$, the block and the rest of the lattice are entangled. The amount of entanglement between these two parts can be measured with the \emph{entanglement entropy}, namely the von Neumann entropy of the reduced density matrix $\rho$, 
\begin{equation}
	S(\rho) \equiv -\tr \left( \rho\log (\rho) \right) = - \sum_{\alpha=1}^{\chi} p_{\alpha}\log (p_{\alpha}).
\end{equation}
This measure vanishes for a product state and its maximum occurs for the flat probability distribution $p_{\alpha} = 1/\chi$, where $S = -\sum_{\alpha} 1/\chi \log (1/\chi) = \log (\chi)$. Therefore we always have $\chi \geq \exp(S)$. In subsequent discussions we will assume for simplicity that
\begin{equation}
	\chi \approx \exp(S).
	\label{V:eq:expS}
\end{equation}

In one spatial dimension, the entanglement entropy of a block of $l$ contiguous sites typically increases with $l$ until $l$ becomes of the order of the correlation length $\xi$ in the system, at which point it saturates to some value $S_{\max}$, whereas it diverges logarithmically at a quantum critical point
\begin{eqnarray}
S(l) &\leq& S_{\max}, ~~~~~~~~~~\mbox{1D non-critical} \label{V:eq:S1D}\\
S(l) &\approx& \frac{c}{6}\log l, ~~~~~~~~~\mbox{1D critical} \label{V:eq:S1Dcrit}
\end{eqnarray}
where $c$ is the central charge of the CFT that describe the infrared limit of the quantum phase transition. In two spatial dimensions, the entanglement entropy of a square block of $l\times l$ sites typically grows proportional to the size of the boundary of the block,
\begin{equation}
	S(l) \approx \alpha l, ~~~~~~~~~~~~~~\mbox{2D}
	\label{V:eq:S2D}
\end{equation}
where $\alpha$ is some constant that depends on the model and grows with $\xi$, and where there might be logarithmic multiplicative corrections in some systems with quasi-long range correlations. More generally, in the ground state of a $D$ dimensional model, the entanglement between a block of $l^D$ sites and the rest of the system scales according to the so-called area law $S(l) \approx \alpha l^{D-1}$, that is, as the size of the boundary of the block, with Eqs. \ref{V:eq:S1D} and \ref{V:eq:S2D} being particular cases of this expression. 

The area law translates into an approximate expression for the effective dimension $\chi^{(\tau)}$. Recall that one site of $\mathcal{L}^{(\tau)}$ accommodates $l = 3^{\tau}$ sites of $\mathcal{L}$ (or $l^2 = 3^{2\tau}$ sites if we had chosen to coarse-grain blocks of $3\times 3$ sites in two spatial dimensions). Eqs. \ref{V:eq:S1D}-\ref{V:eq:S2D} combined with Eq. \ref{V:eq:expS} lead to
\begin{eqnarray}
\chi^{(\tau)} &\leq& \chi_{\max} \approx e^{S_{\max}}  ~~~~~~\mbox{1D non-critical} 
\label{V:eq:Chi1D}\\
\chi^{(\tau)} &\approx& l^{c/6} \approx e^{\tau} ~~~~~~~~~~~~ \mbox{1D critical} 
\label{V:eq:Chi1Dcrit}\\
\chi^{(\tau)} &\approx& e^{\alpha l} \approx e^{e^{\sqrt{\tau}}}~~~~~~~~~~\mbox{2D}
	\label{V:eq:Chi2D}
\end{eqnarray}

Now that we have an expression for the scaling of $\chi^{(\tau)}$, we can analyze both its origin and its implications.
  
\subsection{Accumulation of short-distance degrees of freedom}

From a conceptual viewpoint, the growth of $\chi^{(\tau)}$ reveals an important flaw of the present coarse-graining scheme. Namely, that it fails to remove some of the short-distance degrees of freedom, which remain (and even accumulate) over successive iterations. 

To understand this point, let us assume that two contiguous sites $r,s\in\mathcal{L}$ are in an entangled state, say $(\ket{1_r1_s} + \ket{2_r 2_s})/\sqrt{2}$. We compare two situations, depending on how the blocking in Fig. \ref{V:fig:isometry} affects these sites:

$(i)$ If both sites belong to the same block, then they do not contribute to the entanglement between the block and the rest of the lattice, or what is the same, to the spectrum of the reduced density matrix $\rho$ of the block, and are therefore removed during the coarse-graining. 

($ii$) If the same two sites belong to two adjacent blocks, then they are entangled across the boundary of the blocks, contribute to the spectrum of $\rho$ for each block, and thus will be preserved by the coarse-graining, see Fig. \ref{V:fig:comparison}. 

In other words, short-distance degrees of freedom, entangled at scales smaller than the size of a block, are treated differently depending on whether they are entangled within a block or across the boundary between blocks. In the last case, the coarse-graining transformation fails to remove them.
As a result, the approach cannot generate a proper RG flow. For instance, two Hamiltonians $H_1$ and $H_2$ describing the same phase but differing in short-distance details will remain different under successive iterations of the coarse-graining: some of the short-distance details that distinguish the two models will persist, and two different fixed points of the RG flow will be reached. 

In spite of this shortcoming, is the scheme useful for practical computations? The answer strongly depends on the class of problems under consideration. It is a definitive `yes' for systems in one spatial dimension. White's DMRG \cite{V:White92} (to which this scheme is closely related) revolutionized the computational scene in the early 90's by providing numerical access, with unprecedented accuracy, to ground state energies, local observables and correlators. The immense success of DMRG can be understood from the saturation of $\chi^{(\tau)}$ for non-critical systems, Eq. \ref{V:eq:Chi1D}, since the computational cost is a small power of $\chi^{(\tau)}$. Even critical systems can be studied, in spite of Eq. \ref{V:eq:Chi1Dcrit}, by considering reasonably large chains, of up to thousands of sites, and applying finite-size scaling techniques. Instead, the situation is much grimmer in two spatial dimensions. There the double exponential growth of $\chi^{(\tau)}$ in Eq. \ref{V:eq:Chi2D} implies that only small systems can be addressed (see e.g. the discussion in Ref. \cite{V:Tagliacozzo09}) and there is a real need for alternative approaches.

%%%%%%%%%%%%%%%%%%%%%%%%%%%%%%%%%%%%%%%%%%%%%%%%%%%%%%%%%%%%%%%%%%%%%%%%%%%%%%%%%%%%%%%%%%%%%
%%%%%%%%%%%%%%%%%%%%%%%%%%%%%%%%%%%%%%%%%%%%%%%%%%%%%%%%%%%%%%%%%%%%%%%%%%%%%%%%%%%%%%%%%%%%%

\section{Entanglement Renormalization}
\label{V:sect:ER}

Entanglement renormalization was proposed as a means to consistently remove all short-distance degrees of freedom during coarse-graining. It leads to a significant reduction in computational costs, making scalable calculations feasible even in two dimensions. More important to us, it produces a RG flow with the expected properties.

%%%%%%%%%%%%%%%%%%%%%%%%%%%%%%%%%%%%%%%%%%%
\begin{figure}
\begin{center}
\includegraphics[width=10cm]{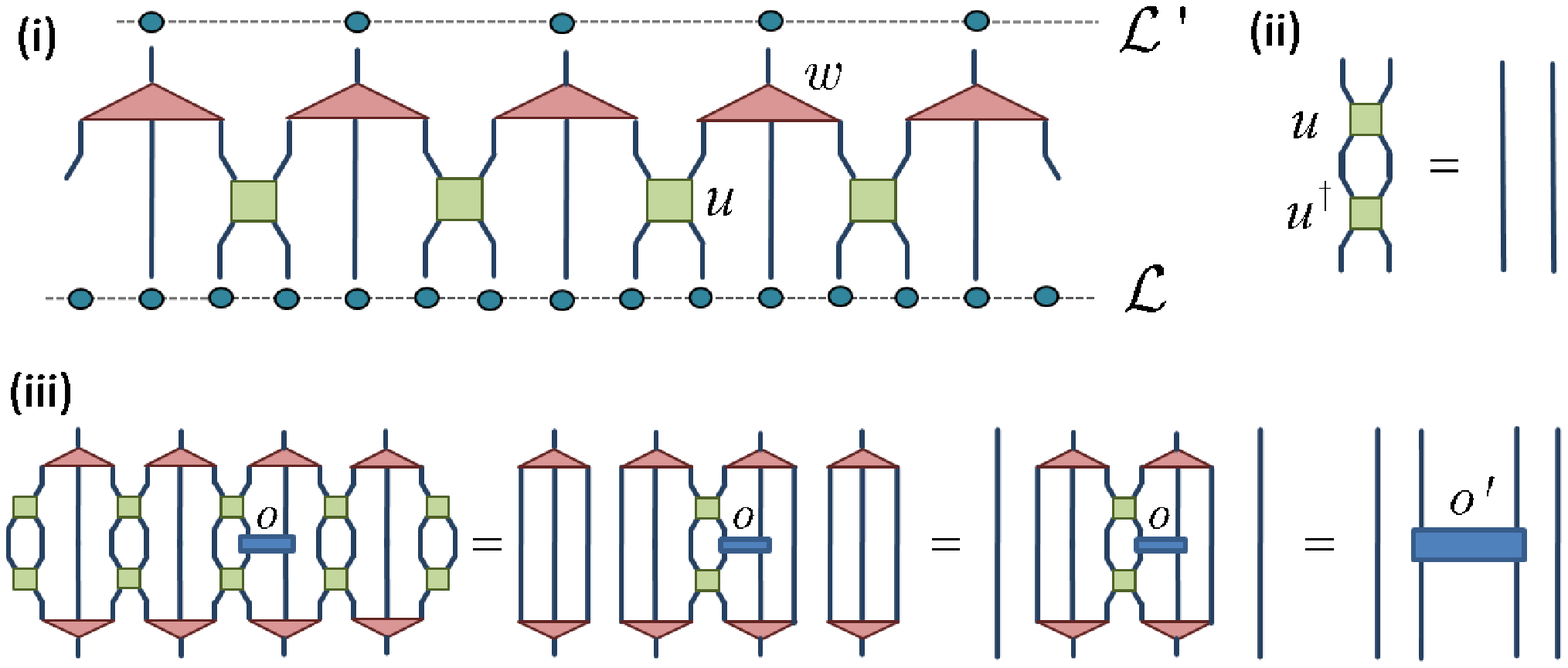}
\end{center}
\caption{ 
(i) Coarse-graining transformation with disentanglers $u$, acting on two contiguous sites of $\mathcal{L}$ across the boundary of two blocks, followed by isometries $w$, which map a block of three sites into a single site of the coarse-grained lattice $\mathcal{L}'$. (ii) Graphical representation of $uu^{\dagger} = I_{\mathbb{V}^{\otimes 2}}$. (iii) A local operator $o$ supported on two contiguous sites of $\mathcal{L}$ is mapped into an effective operator $o'$ on two contiguous sites of $\mathcal{L}'$. The linear transformation $o\rightarrow o' = \mathcal{A}(o)$ is described by a superoperator (i.e., a linear map in the space of operators) referred to as the ascending superoperator $\mathcal{A}$.
}
\label{V:fig:disentangler}
\end{figure}
%%%%%%%%%%%%%%%%%%%%%%%%%%%%%%%%%%%%%%%%%%%

\subsection{Disentanglers}

Let us modify the previous coarse-graining scheme as follows. First we apply a unitary transformation $u$, a \emph{disentangler}, on pairs of contiguous sites of $\mathcal{L}$,
\begin{equation}
	u: \mathbb{V}^{\otimes 2} \longmapsto \mathbb{V}^{\otimes 2},~~~~~~u^{\dagger}u = u u^{\dagger} = I_{\mathbb{V}^{\otimes 2}},
	\label{V:eq:u}
\end{equation}
where each site belongs to one of two adjacent blocks, so that $u$ acts across the boundary of the two blocks, see Fig. \ref{V:fig:disentangler}. Then, as before, we use isometries $w$ to map the blocks of $3$ sites of $\mathcal{L}$ into single sites of the lattice $\mathcal{L}'$\footnote{The distinction between disentanglers $u$ and isometries $w$ is useful for pedagogical purposes. However, more generally one can consider isometric transformations $v:\mathbb{V}^{\otimes n_1} \mapsto \mathbb{V}^{\otimes n_2}$, with $v^{\dagger}v = I_{\mathbb{V}^{\otimes n_1}}$, $n_1 \leq n_2$, that both disentangle the ground state and coarse-grain the lattice.}. Let $U\equiv u^{\otimes N/3}$ be a unitary transformation that contains disentanglers $u$ acting on the boundaries of all the blocks. Then the effective Hamiltonian
in $\mathcal{L}'$ reads
\begin{equation}
	H' = \left(W^{\dagger}U^{\dagger}\right) H \left(U W\right).
\end{equation}

\begin{figure}
\begin{center}
\includegraphics[width=10cm]{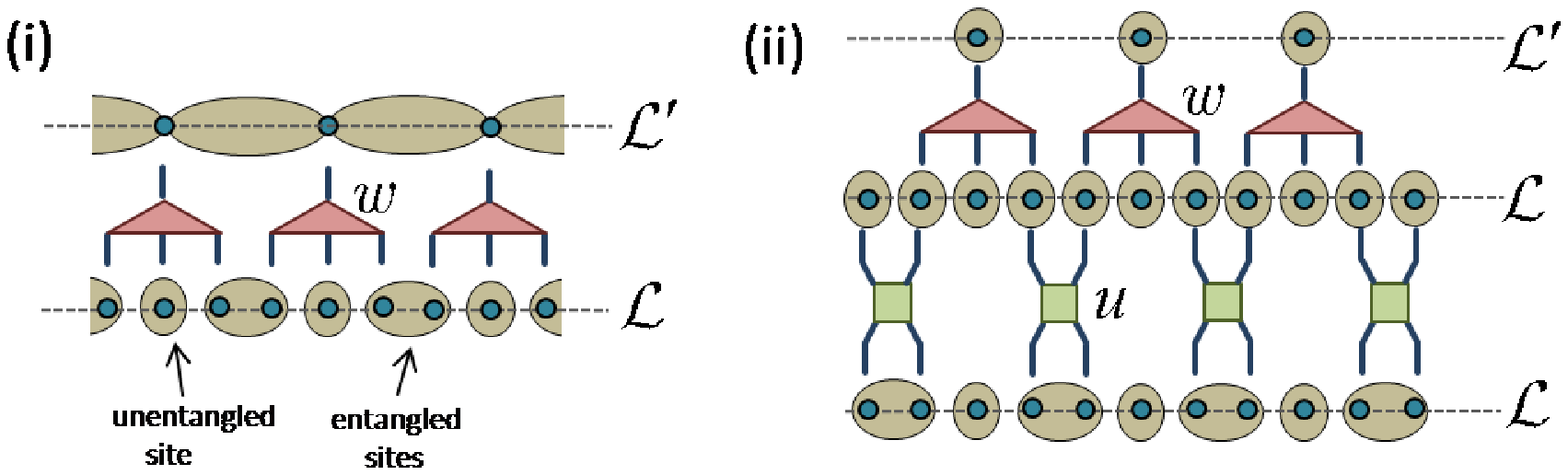}
\end{center}
\caption{To illustrate the role of disentanglers $u$, a simple example is considered where the ground state $\ket{\GS}$ of a one-dimensional lattice $\mathcal{L}$ factorizes into the product of entangled states $(\ket{1_r 1_s} + \ket{2_r 2_s})/\sqrt{2}$ involving only two nearest neighbor sites $r,s \in \mathcal{L}$ (on both sides of the boundaries between blocks) and states of single sites (in the interior of each block). (i) When only isometries are used, entanglement across the boundary of the blocks is preserved in $\mathcal{L}'$. (ii) By using disentanglers $u$ such that transform the state $(\ket{1_r 1_s} + \ket{2_r 2_s})/\sqrt{2}$ of the two boundary sites $r,s\in\mathcal{L}$ into an unentangled state, e.g.  $\ket{1_r1_s}$, entanglement across the boundary of the blocks can be removed before the isometries are applied, and the ground state of $\mathcal{L}'$ has no entanglement.}
\label{V:fig:comparison}
\end{figure}

In order to guarantee the preservation of ground state properties, the isometries $w$ are chosen according to White's rule, namely such that they retain the support of the reduced density matrix $\tilde{\rho}$ on the block, 
\begin{equation}
	\tilde{\rho} = \mbox{tr}_{\mathcal{L}/block} \left( U^{\dagger} \proj{\GS} U\right).
\end{equation}
Notice, however, that $\ket{\GS}$ has been transformed according to the disentanglers before the density matrix $\tilde{\rho}$ is computed. By properly choosing the disentanglers $u$, the modified ground state $\ket{\tilde{\Psi}_{\gs}} = U \ket{\GS}$ will have less entanglement than $\ket{\GS}$. Specifically, the disentanglers are capable of removing short-range entanglement across the boundary of the blocks, see Fig. \ref{V:fig:comparison}.

%%%%%%%%%%%%%%%%%%%%%%%%%%%%%%%%%%%%%%%%%%%
%%%%%%%%%%%%%%%%%%%%%%%%%%%%%%%%%%%%%%%%%%%

\begin{figure}
\begin{center}
\includegraphics[width=11cm]{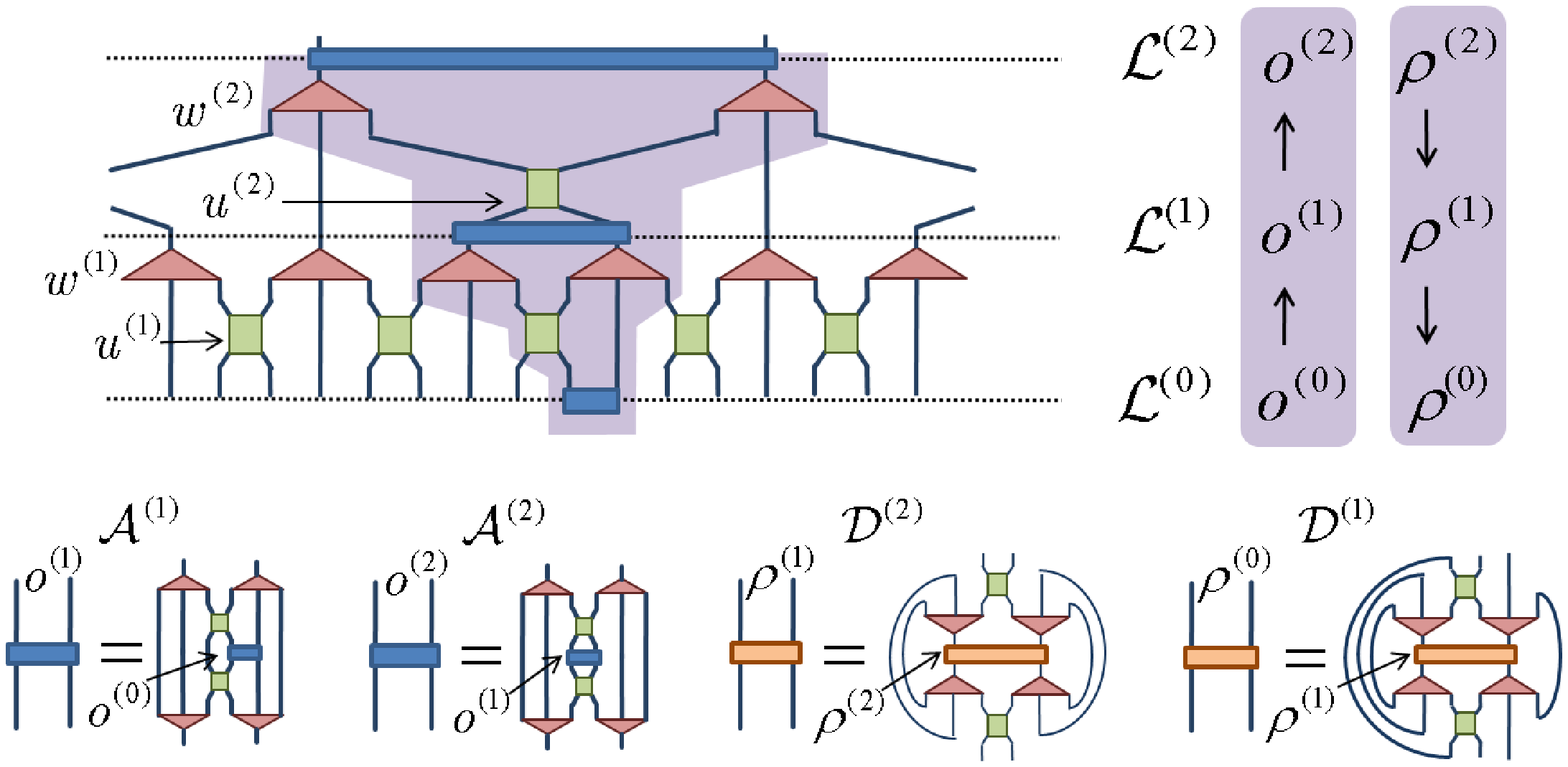}
\end{center}
\caption{Diagramatic representation of two iterations of the coarse-graining transformation, producing a sequence of increasingly coarse-grained lattices $\mathcal{L}^{(0)}$, $\mathcal{L}^{(1)}$, and $\mathcal{L}^{(2)}$. At each iteration $u^{(\tau)}$ is chosen as to remove short-range entanglement and $w^{(\tau)}$ follows White's rule. The ascending superoperator $\mathcal{A}$ maps $o^{(\tau)}$ into $o^{(\tau+1)}$ while the descending superoperator $\mathcal{D}$ maps $\rho^{(\tau+1)}$ into $\rho^{(\tau)}$. 
}
\label{V:fig:causal}
\end{figure}

By iterating the transformation we can once more build a sequence of increasingly coarse-grained models (Fig. \ref{V:fig:causal})
\begin{eqnarray}
	(\mathcal{L}^{(0)}, H^{(0)}) \longrightarrow (\mathcal{L}^{(1)}, H^{(1)}) \longrightarrow (\mathcal{L}^{(2)}, H^{(2)}) \longrightarrow \cdots,
	\label{V:eq:sequence}
\end{eqnarray}
As evidenced by abundant numerical and analytical results (Refs. \cite{V:Vidal07}-\cite{V:Aguado09}), an important consequence of the use of disentanglers is that the dimension $\chi^{(\tau)}$ of the effective sites no longer needs to increase with $\tau$ in order to (approximately, but accurately) fulfill White's rule.
This is the case in one spatial dimension, both for critical and non-critical systems, cf. Eqs. \ref{V:eq:S1D}-\ref{V:eq:S1Dcrit}, as well as in two spatial dimensions, provided the area law of Eq. \ref{V:eq:S2D} does not have logarithmic corrections.
 
%%%%%%%%%%%%%%%%%%%%%%%%%%%%%%%%%%%%%%%%%%%
%%%%%%%%%%%%%%%%%%%%%%%%%%%%%%%%%%%%%%%%%%%

\subsection{Ascending and descending superoperators}

An operator $o$ supported on two contiguous sites of $\mathcal{L}$ is now mapped into an operator $o'$ supported on two contiguous sites of $\mathcal{L'}$ by means of the ascending superoperator $\mathcal{A}$ (Fig. \ref{V:fig:disentangler}),
\begin{equation}
	o \rightarrow o' = \left( (w^{\dagger\otimes 2})(u^{\dagger} )\right) o \left( u (w ^{\otimes 2})\right) \equiv \mathcal{A}(\rho).
	\label{V:eq:ascending}
	\end{equation}
Notice that locality is again preserved under the coarse-graining transformation, but in this case it results from a compromise between two opposing forces: disentanglers $u$ expand the support of local operators, while isometries $w$ compress it. The balance corresponds to two-site supports, to which both smaller and larger supports tend under coarse-graining. This is why two-site supports play a dominant role in the present discussions. In the case of a sequence of coarse-graining transformations (Fig. \ref{V:fig:causal}) we can investigate how the two-site operator $o^{(0)}\equiv o$ changes as we increase the scale of observation:
\begin{equation}
	o^{(0)} \stackrel{\mathcal{A}^{(1)}}{\longrightarrow} o^{(1)} \stackrel{\mathcal{A}^{(2)}}{\longrightarrow} o^{(2)} \longrightarrow \cdots
\end{equation}
On the other hand, the descending superoperator $\mathcal{D}$,
\begin{equation}
	\rho' \rightarrow \rho = \mbox{tr}_{s_1s_2} \left(  (w ^{\otimes 2}) u \right)\rho' \left( (w^{\dagger\otimes 2})(u^{\dagger} )\right),
\end{equation}
where $\rho$ and $\rho'$ are supported on the same two sites as $o$ and $o'$ in Eq. \ref{V:eq:ascending} and the partial trace eliminates two superfluous sites $s_1,s_2\in\mathcal{L}$, produces a fine-grained reduced density matrix $\rho$ from a coarse-grained reduced density matrix $\rho'$, such that
\begin{equation}
	\tr(\rho o) = \tr(\rho' o'),
\end{equation}
for all possible two-site operators $o$\footnote{In other words, the descending superoperator $\mathcal{D}$ is dual to the ascending superoperator $\mathcal{A}$, $\tr(\rho' \mathcal{A}(o)) = \tr(\mathcal{D}(\rho') o)$ for all $\rho'$ and $o$.}.
By iteration, the descending superoperator allows us to obtain a sequence of two-site density matrices 
\begin{equation}
	\cdots \longrightarrow \rho^{(2)} 
	\stackrel{\mathcal{D}^{(2)}}{\longrightarrow} \rho^{(1)}
	\stackrel{\mathcal{D}^{(1)}}{\longrightarrow} \rho^{(0)} 
\end{equation}
flowing from coarser to finer lattices and therefore monitor how the ground state two-site density matrix changes with the observation scale.

\subsection{Multi-scale Entanglement Renormalization Ansatz}

\begin{figure}
\begin{center}
\includegraphics[width=8cm]{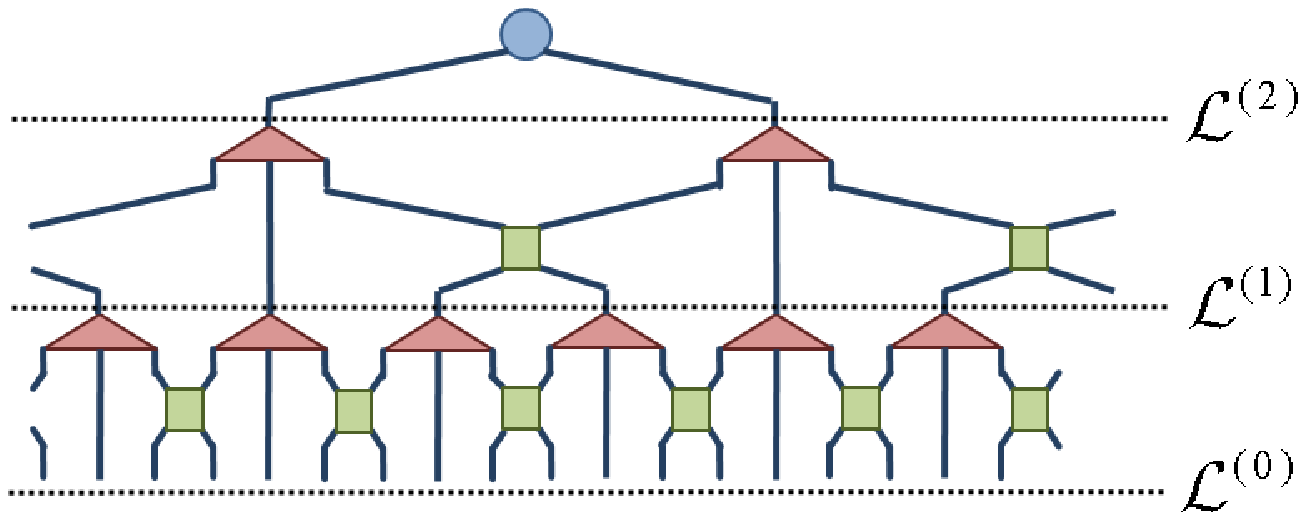}
\end{center}
\caption{Multi-scale entanglement renormalization ansatz made of disentanglers and isometries corresponding to two iterations of the coarse-graining transformation in Fig. \ref{V:fig:disentangler}. Notice the periodic boundary conditions. 
}
\label{V:fig:MERA}
\end{figure}

The \emph{multi-scale entanglement renormalization ansatz} (MERA) is a variational ansatz for pure states $\ket{\Psi}\in\mathbb{V}^{\otimes N}$ of $\mathcal{L}$  that results in a natural way from the above coarse-graining transformation. It consists of a tensor network that collects all the disentanglers and isometries used to produce the sequence of coarse-grained lattices in Eq. \ref{V:eq:sequence}, see Fig. \ref{V:fig:MERA}. 

In the case of a translation invariant system, we need to specify one disentangler $u^{(\tau)}$ and one isometry $w^{(\tau)}$ per layer, and these depend on $O(\chi^q)$ parameters, where $q=4$ in the one-dimensional scheme discussed here. Since there are $O(\log(N))$ layers of tensors, the MERA depends on $O(\chi^q \log (N))$ parameters, a result valid also in two spatial dimensions for an appropriate value of $q$. 

Therefore, the MERA is an efficient representation of certain states of lattice $\mathcal{L}$. The interest in this ansatz resides in that abundant numerical and analytical results demonstrate that it can be used to accurately approximate the ground state $\ket{\GS}$ of a local Hamiltonian $H$ in large (and even infinite) lattice systems in one and two spatial dimensions. Even at criticality the error in ground state expected values, Eq. \ref{V:eq:oGS}, decays exponentially with the refinement parameter $\chi$. While it is still unclear exactly what Hamiltonians have ground states that can be well approximated, a MERA with a fixed value of $\chi$ naturally reproduces a scaling of entanglement compatible with Eqs. \ref{V:eq:S1D}-\ref{V:eq:S2D}, suggesting that it might be able to accurately approximate ground states whose block entanglement fulfills the area law. 

Given the ground state $\ket{\GS}$ of a local Hamiltonian $H$, we have provided an intuitive description of the role played by the tensors that form the MERA. Namely, a disentangler $u$ `removes short-range entanglement from $\ket{\GS}$', while an isometry $w$ `coarse-grains a block of sites so as to preserve the support of the ground state reduced density matrix'. These descriptions could perhaps allow us to compute all $u$'s and $w$'s of the MERA for $\ket{\GS}$ if we already knew $\ket{\GS}$. Algorithmically, of course, it is absurd to assume that we already know $\ket{\GS}$ if our goal is to compute $\ket{\GS}$. In practical MERA calculations, an approximation to $\ket{\GS}$ is obtained by minimizing the expected value $\bra{\Psi} H \ket{\Psi}$, where $\ket{\Psi}$ is constrained to be a MERA. This is done starting with random $u$'s and $w$'s and iteratively optimizing them so as to reduce the energy, with a cost that scales as $O(\chi^{q'}\log (N))$ per optimization step ($q'=7$ in the present scheme). We refer to the literature for further details. 

%%%%%%%%%%%%%%%%%%%%%%%%%%%%%%%%%%%%%%%%%%%
%%%%%%%%%%%%%%%%%%%%%%%%%%%%%%%%%%%%%%%%%%%

\section{The Renormalization Group picture}
\label{V:sect:RG}

From now on we consider an infinite lattice $\mathcal{L}$ and a Hamiltonian $H$ that is the sum of nearest neighbor terms 
\begin{equation}
H = \sum_{s} h(s,s+1),~~~~~~h(s,s+1) \equiv h,	
\end{equation}
where the two-site operator $h$ is the same on all pairs of nearest neighbor sites, and thus $h$ completely characterizes $H$. We assume that disentanglers and isometries have been properly optimized, e.g. by minimizing the expected value $\bra{\Psi}H\ket{\Psi}$, and concentrate on analyzing the properties of the resulting coarse-graining from the perspective of the renormalization group.

\subsection{A real space RG map}

Let us introduce the average ascending superoperator $\bar{\mathcal{A}}^h$,
\begin{equation}
	o' = \bar{\mathcal{A}}^h(o) \equiv \frac{\mathcal{A}^{h}_{L}(o) + \mathcal{A}^{h}_{C}(o) + \mathcal{A}^{h}_{R}(o)}{3},
\end{equation}
that averages over the three inequivalent ways (left, center and right) in which $o$ can be coarse-grained, and where the superscript $h$ highlights the dependence on the Hamiltonian $H$---recall that the ascending superoperator is built with the disentanglers and isometries corresponding to the ground state $\ket{\GS}$ of $H$. Under coarse-graining, $H$ is mapped into an effective Hamiltonian $H'$,
\begin{equation}
H' = 3 \sum_{s} h'(s,s+1),~~~~~~h'(s,s+1) \equiv h',	
\end{equation}
where the constant two-site operator $h'$ is obtained from $h$ by the RG map  $\mathcal{R}$,
\begin{equation}
	h' =  \mathcal{R}(h) \equiv \bar{\mathcal{A}}^h(h).
\end{equation}
Notice that, by construction, $\mathcal{R}(h)$ is a non-linear function of $h$. 

Given the two-site interaction term $h^{(0)}\equiv h$ of the initial Hamiltonian $H^{(0)}\equiv H$, we can now build a sequence of two-site interactions
\begin{equation}
	h^{(0)}
\stackrel{\mathcal{R}}{\longrightarrow} h^{(1)} \stackrel{\mathcal{R}}{\longrightarrow} h^{(2)} \stackrel{\mathcal{R}}{\longrightarrow}    \cdots
\end{equation}
where $h^{(\tau)}$ characterizes the Hamiltonian $H^{(\tau)} = 3^{\tau} \sum_s h^{(\tau)}(s,s+1)$ of the coarse-grained lattice $\mathcal{L}^{(\tau)}$. This defines a \emph{discrete} RG flow in the ($\chi^4$-dimensional) space of possible two-site interactions that we can use to study how the Hamiltonian of the system changes with the observation scale.

%%%%%%%%%%%%%%%%%%%%%%%%%%%%%%%%%%%%%%%%%%%
\begin{figure}
\begin{center}
\includegraphics[width=11cm]{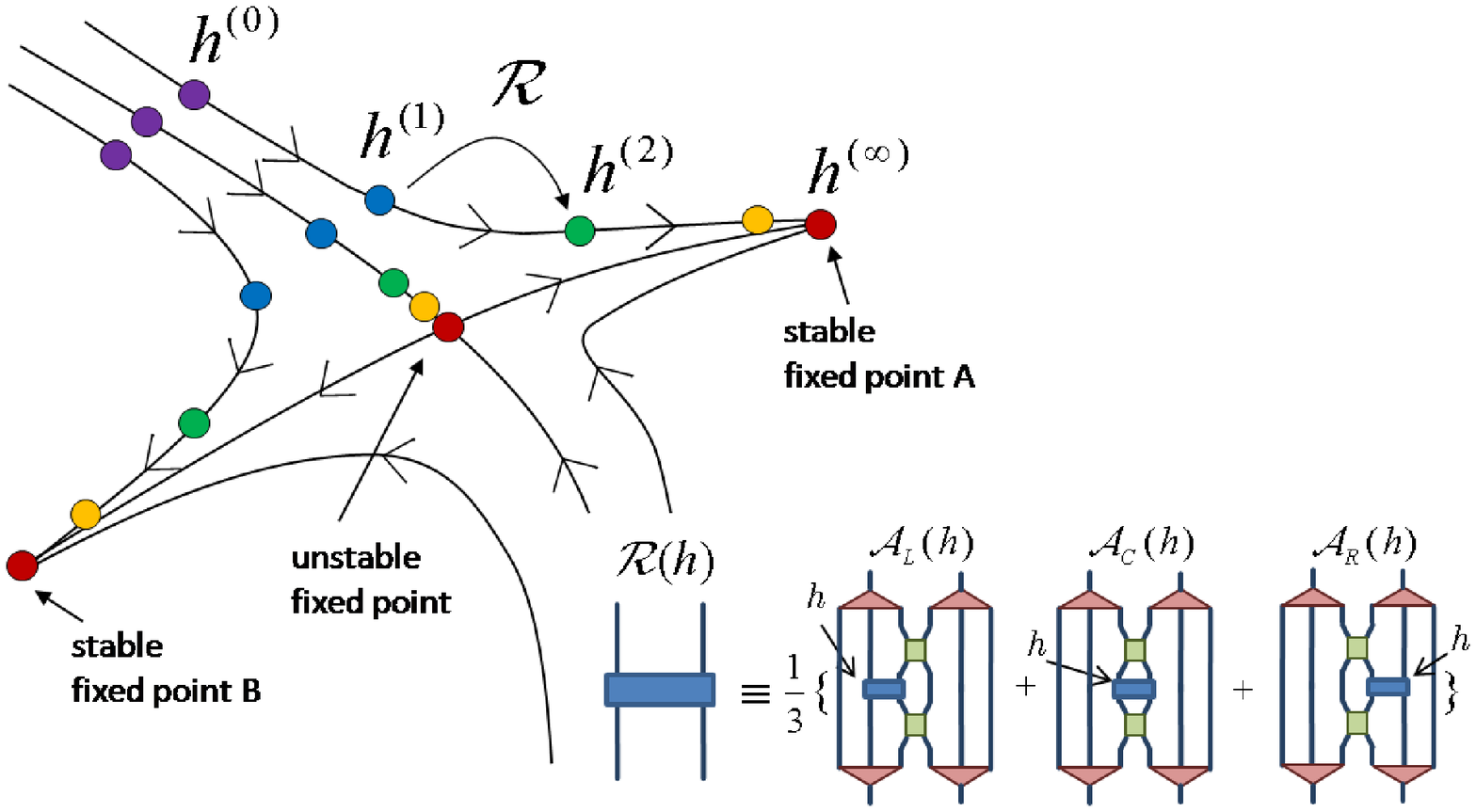}
\end{center}
\caption{The RG map $\mathcal{R}(h)$ defines a RG flow in the space of two-site interactions $h$. The scaling superoperator $\mathcal{S}(o)$, Eq. \ref{V:eq:S}, transforms two-site operators $o$ at the fixed points of this RG flow.
}
\label{V:fig:RG}
\end{figure}
%%%%%%%%%%%%%%%%%%%%%%%%%%%%%%%%%%%%%%%%%%%

\subsection{Properties of the RG map}

Let us briefly discuss a few properties of the RG map $\mathcal{R}(h)$:

($i$) \emph{Proper RG flow}.--- Abundant numerical evidence suggests that properly chosen disentanglers indeed succeed at removing \emph{all} short-distance degrees of freedom. As a result, if two Hamiltonians $H_1$ and $H_2$ correspond to the same phase, multiple applications of the RG map take $h_1^{(0)}$ and $h_2^{(0)}$ into the same fixed-point interaction $h^{*}$, that is $h_1^{*}=h_2^{*}$ (up to trivial changes of local basis). 

($ii$) \emph{Nearest neighbor interactions}.--- As mentioned earlier, the present coarse-graining transformation does not generate long-range interactions starting from a short-ranged Hamiltonian (in contrast e.g. with momentum-space RG methods). In addition, a Hamiltonian containing arbitrary short-range interactions (well beyond nearest neighbors) can be reduced, after a few iterations, to a Hamiltonian with only nearest neighbor interactions\footnote{A coarse-graining step with trivial disentanglers is required (once) in order to eliminate some of the next-to-nearest neighbor interactions.}. Therefore the two-site RG map $\mathcal{R}(h)$ can be used to study arbitrary phases with short-range interactions. In two spatial dimensions, the analogous construction leads to an RG flow for four-site Hamiltonians.

($iii$) \emph{Unbiased RG map}.--- Notice that the space $\mathbb{V}'$ for an effective site of $\mathcal{L}'$ is not chosen a priori on the basis of heuristic arguments, nor is some specific form of the effective Hamiltonian $H'$ (with a few free parameters to be fixed) imposed. The approach simply searches, through an energy minimization, the subspace $\mathbb{V}^{'\otimes N/3}$ of the total Hilbert space $\mathbb{V}^{\otimes N}$ that best approximates the low energy subspace of $H$. The only hypothesis that is made concerns the structure of the subspace $\mathbb{V}^{'\otimes N/3}$, assumed to factorize as a tensor product of spaces $\mathbb{V}'$ that are related to the original factorization $\mathbb{V}^{\otimes N}$ through the disentanglers $u$ and isometries $w$. Numerical evidence (provided e.g. by the study of critical phenomena in the next section) suggest that the low energy subspace of many local Hamiltonians of interest indeed has this structure.

\subsection{Fixed points of entanglement renormalization}

Following the RG program, our next step is to study the fixed points of the flow generated by the RG map $\mathcal{R}$, corresponding to Hamiltonians that do not change when we modify the scale of observation. That is, we are interested in models with two-site interactions $h^{\star}$ such that 
\begin{equation}
	\mathcal{R}(h^{\star}) = \lambda h^{\star} ~~~~~~~~~~~\mbox{scale invariance}
	\label{V:eq:fixedpoint}
\end{equation}
where $\lambda \equiv 3^{-\Delta_{h^{\star}}}$ is a proportionality constant to be discussed later on.

A fixed point is characterized by one disentangler $u$ and one isometry $w$ that are repeated at all length scales, i.e. $u^{(\tau)} = u$ and $w^{(\tau)} = w$ for all $\tau$. Thus, the \emph{scale invariant} MERA, specified by this pair $(u,w)$, depends on just $O(\chi^4)$ parameters and offers an extremely compact description of fixed-point ground states of an infinite lattice. It can be obtained with an algorithm whose cost formally scales as $O(\chi^7)$ \cite{V:Pfeifer09}.

Since a finite correlation length $\xi$ shrinks under coarse-graining, scale invariance requires that $\xi=0$ or $\xi=\infty$, corresponding to non-critical and critical fixed points. The present approach leads to a natural characterization of RG fixed points in terms of the entanglement of their ground states and the MERA representation:

($i$) Non-critical fixed-point ground states ($\xi=0$) may be unentangled or entangled. Symmetry-breaking phases usually have an unentangled (or product) fixed-point ground state, represented with a trivial scale invariant MERA with $\chi=1$, whereas the fixed-point ground state in topologically ordered phases is entangled. The results of \cite{V:Aguado08, V:Koenig08} strongly suggest that a scale invariant MERA with finite $\chi$ can exactly represent such entangled ground states.

($ii$) Critical fixed-point ground states ($\xi=\infty$) are always highly entangled. An exact MERA representation requires an infinite dimension $\chi$, as will be discussed in Sect. \ref{V:sect:QuCrit}, but we will also see that accurate estimates of critical properties can be obtained from a scale invariant MERA with finite $\chi$.

In order to further characterize a fixed point, one usually linearizes $\mathcal{R}$ at $h=h^{\star}$ by considering its derivative with respect to a small perturbation $\epsilon o$,
\begin{equation}
	\lim_{\epsilon \rightarrow 0} \frac{\mathcal{R}(h^{\star}+\epsilon o) - \mathcal{R}(h^{\star})}{\epsilon} = 	\lim_{\epsilon \rightarrow 0} \frac{\bar{\mathcal{A}}^{h^{\star}+\epsilon o}(h^{\star}+\epsilon o) - \bar{\mathcal{A}}^{h^{\star}}(h^{\star})}{\epsilon}.
\end{equation}
We conjecture\footnote{This conjecture would be true if $\lim_{\epsilon\rightarrow 0} (\bar{A}^{h^{\star}+\epsilon o}(h^{\star})-\bar{A}^{h^{\star}}(h^{\star}))/\epsilon$=0, which at a critical point is both plausible and compatible with numerical tests.} that this derivative is given by $\bar{\mathcal{A}}^{h^{\star}}(o)$, which dictates how two-site operators $o$ transform under coarse-graining at the fixed point $h=h^{\star}$, and which we will call \emph{scaling superoperator} $\mathcal{S}$,
\begin{equation}
	\mathcal{S}(o) \equiv \bar{\mathcal{A}}^{h^{\star}}(o).
	\label{V:eq:Scaling}
\end{equation}
Recall that one can distinguish between stable fixed points, where \emph{any} perturbed interaction $h^{\star}+\epsilon o$ will flow back to $h^{\star}$; and unstable fixed points, where relevant perturbations $o$ exists such that $h^{\star}+\epsilon o$ will flow away from $h^{\star}$. It may also be worth recalling that the notion of stability is relative to what perturbations $o$ are available, and thus may depend on symmetry considerations. As illustrated in the next section for critical fixed points, from the scaling superoperator $\mathcal{S}$ one can identify all relevant perturbations of a fixed point and therefore analyze its stability.

%%%%%%%%%%%%%%%%%%%%%%%%%%%%%%%%%%%%%%%%%%%%%%%%%%%%%%%%%%%%%%%%%%%%%%%%%%%%%%%%%%%%%%%%%%%%%
%%%%%%%%%%%%%%%%%%%%%%%%%%%%%%%%%%%%%%%%%%%%%%%%%%%%%%%%%%%%%%%%%%%%%%%%%%%%%%%%%%%%%%%%%%%%%

\section{Quantum criticality} 
\label{V:sect:QuCrit}

In this section we apply the scale invariant MERA to the study of critical fixed points of the RG flow, that is to quantum critical points. In this way we describe an important application of entanglement renormalization, while also demonstrating the validity of the approach. For concreteness, we analyze the quantum Ising chain with critical transverse magnetic field,
\begin{equation}
	H = \sum_{s} \sigma_x(s)\otimes\sigma_x(s+1) + \sum_{s} \sigma_z(s),
\end{equation}
whose well-known critical properties are described by a (1+1) conformal field theory (CFT) \cite{V:Henkel99}.

\subsection{Scaling operators and critical exponents}

%%%%%%%%%%%%%%%%%%%%%%%%%%%%%%%%%%%%%%%%%%%
\begin{figure}
\begin{center}
\includegraphics[width=10cm]{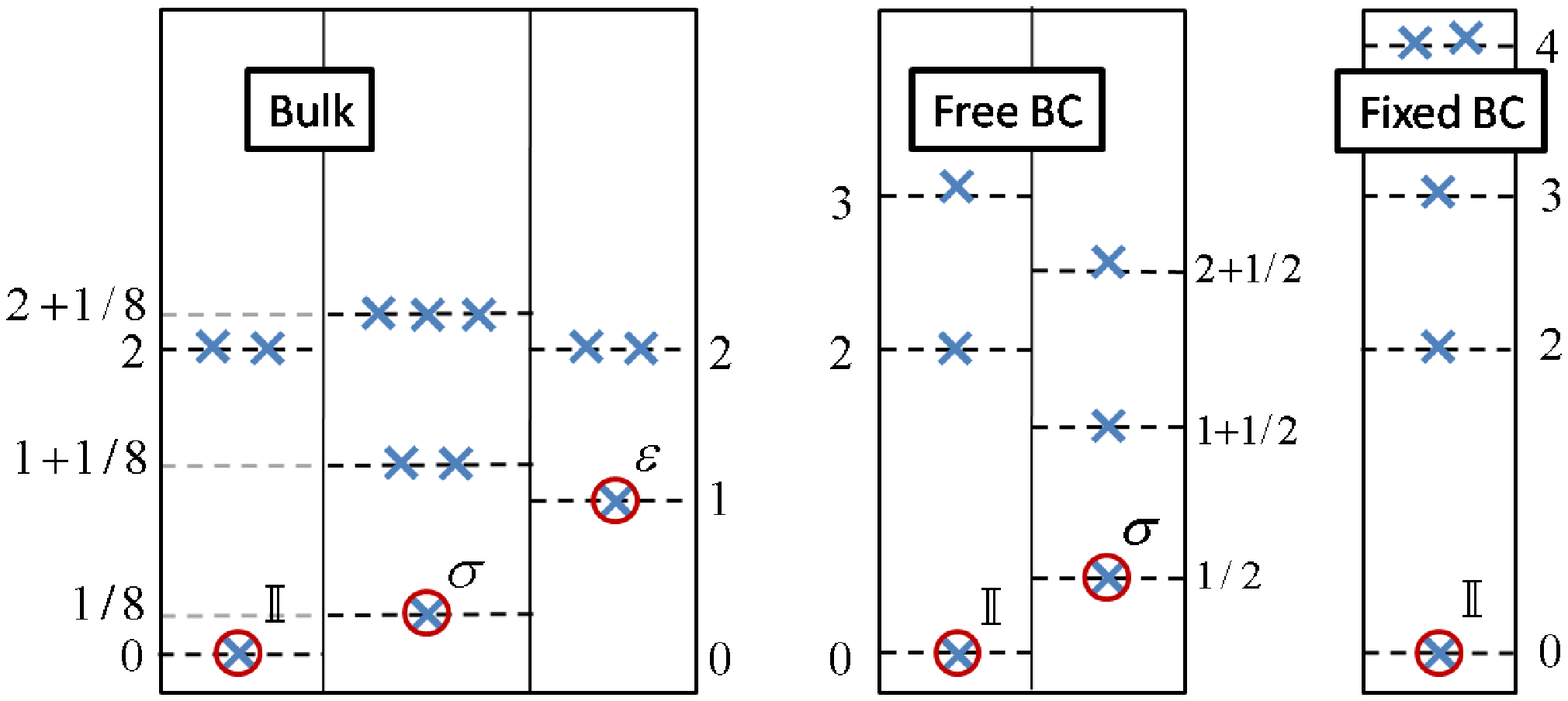}
\end{center}
\caption{Scaling dimensions of the quantum critical Ising model extracted from the bulk and boundary scaling superoperators of a $\chi=16$ MERA.
}
\label{V:fig:ScaleDim}
\end{figure}
%%%%%%%%%%%%%%%%%%%%%%%%%%%%%%%%%%%%%%%%%%%

Most properties of a quantum critical point can be extracted from the spectral decomposition of the (linearized) RG map at the corresponding fixed point of the RG flow, see e.g. \cite{V:Cardy02}. In the present case, the spectral decomposition of the scaling superoperator $\mathcal{S}$ \cite{V:Giovannetti08,V:Pfeifer09},
%\footnote{We assume, for simplicity, that there are no Jordan block and therefore $\mathcal{S}$ has $\chi^2$ "eigenvectors" $\phi_{\alpha}$.}
\begin{equation}
	\mathcal{S}(\bullet) = \sum_{\alpha} \lambda_{\alpha}\phi_{\alpha} \tr(\hat{\phi}_{\alpha}\bullet),~~~~~~~ \tr(\hat{\phi}_{\alpha} \phi_{\beta})=\delta_{\alpha\beta}, \label{V:eq:spectrum}
\end{equation}
readily provides us with the scaling operators of the lattice model, namely those operators $\phi_{\alpha}$ on two contiguous sites that transform into themselves under the coarse-graining transformation,
\begin{equation}
	\mathcal{S}(\phi_{\alpha}) = \lambda_{\alpha} \phi_{\alpha},~~~~~~~\lambda_{\alpha} \equiv 3^{-\Delta_{\alpha}},
\label{V:eq:S}
\end{equation}
as well as their scaling dimensions $\Delta_{\alpha} = -\log_3(\lambda_{\alpha})$. An example of scaling operator is the identity operator $\phi_{\mathbb{I}} \equiv \mathbb{I}$,
\begin{equation}
	\mathcal{S}(\mathbb{I}) = \mathbb{I},~~~~~~~~~\Delta_{\mathbb{I}}=0,
\end{equation}
as can be seen by using $u^{\dagger}u=I$ and $w^{\dagger}w = I $ in Eq. \ref{V:eq:ascending}. Its dual operator $\hat{\phi}_{\mathbb{I}}$ in Eq. \ref{V:eq:spectrum} is the ground state reduced density matrix $\hat{\phi}_{\mathbb{I}} \equiv \hat{\rho}$ for a block of two contiguous sites. A second example of scaling operator is of course the two-site fixed-point interaction $h^*$, see Eq. \ref{V:eq:fixedpoint}, with scaling dimension $\Delta_{h^{\star}}=2$. 

%%%%%%%%%%%%%%%%%%%%%%%%%%%%%%%%%%%%%%%%%%%%%%%%%%%%%%%%%%%%%%%%%%%%%%%%%%%
\begin{table}
\caption{Scaling dimensions of the critical quantum Ising chain.}
\begin{tabular}{||l|l|c||c|c|c||}
  \hline
 $~~~~~\Delta^{\mbox{\tiny CFT}}$ &  $~\Delta^{\mbox{\tiny MERA}}_{\mbox{\tiny $\chi=16$}}~$ & error & 
 $~~~\Delta^{\mbox{\tiny CFT}}$ & $~\Delta^{\mbox{\tiny MERA}}_{\mbox{\tiny $\chi=16$}}~$  & error\\ \hline
     ($\mathbb{I}\,$) 0  &  0      & --          &  2      &  1.99956  & 0.022$\%$ \\ 
     ($\sigma$) 0.125   &  0.124997& 0.003$\%$   &  2      &  1.99985  & 0.007$\%$ \\  
     ($\epsilon\,$) 1   &  0.99993 & 0.007$\%$   &  2      &  1.99994  & 0.003$\%$ \\ 
     $~~~~~$1.125       &  1.12495 & 0.005$\%$   &  2      &  2.00057  & 0.03$\%$  \\ 
     $~~~~~$1.125       &  1.12499 & 0.001$\%$   &         &           & \\
  \hline
  \end{tabular} \nonumber
\label{V:table:scaling}
\end{table}
%%%%%%%%%%%%%%%%%%%%%%%%%%%%%%%%%%%%%%%%%%%%%%%%%%%%%%%%%%%%%%%%%%%%%%%%%%%

Fig. \ref{V:fig:ScaleDim} shows the 12 smallest scaling dimensions for the Ising model, obtained with a scale invariant MERA with $\chi=16$. As expected, they appear organized in the \emph{conformal towers} of three scaling operators known as the \emph{identity} $\mathbb{I}$, the \emph{spin} $\sigma$, and the \emph{energy} $\epsilon$, which correspond to the \emph{primary fields} of the Ising CFT \cite{V:Henkel99}. Recall that all critical exponents of a model can be obtained from the scaling dimensions of its primary fields. For instance, for the Ising model the exponents $\nu$ and $\eta$ are $\nu = 2\Delta_{\sigma}$ and $\eta = \frac{1}{2-\Delta_{\epsilon}}$, and the \emph{scaling laws} express the critical exponents $\alpha,\beta,\gamma, \delta$ in terms of $\nu$ and $\eta$. We can also identify e.g. all the relevant perturbations of the model, corresponding to scaling dimension $\Delta < 2$, as well as its marginal perturbations, with $\Delta = 2$, which are listed in Table \ref{V:table:scaling}.

%%%%%%%%%%%%%%%%%%%%%%%%%%%%%%%%%%%%%%%%%%%
\begin{figure}
\begin{center}
\includegraphics[width=10cm]{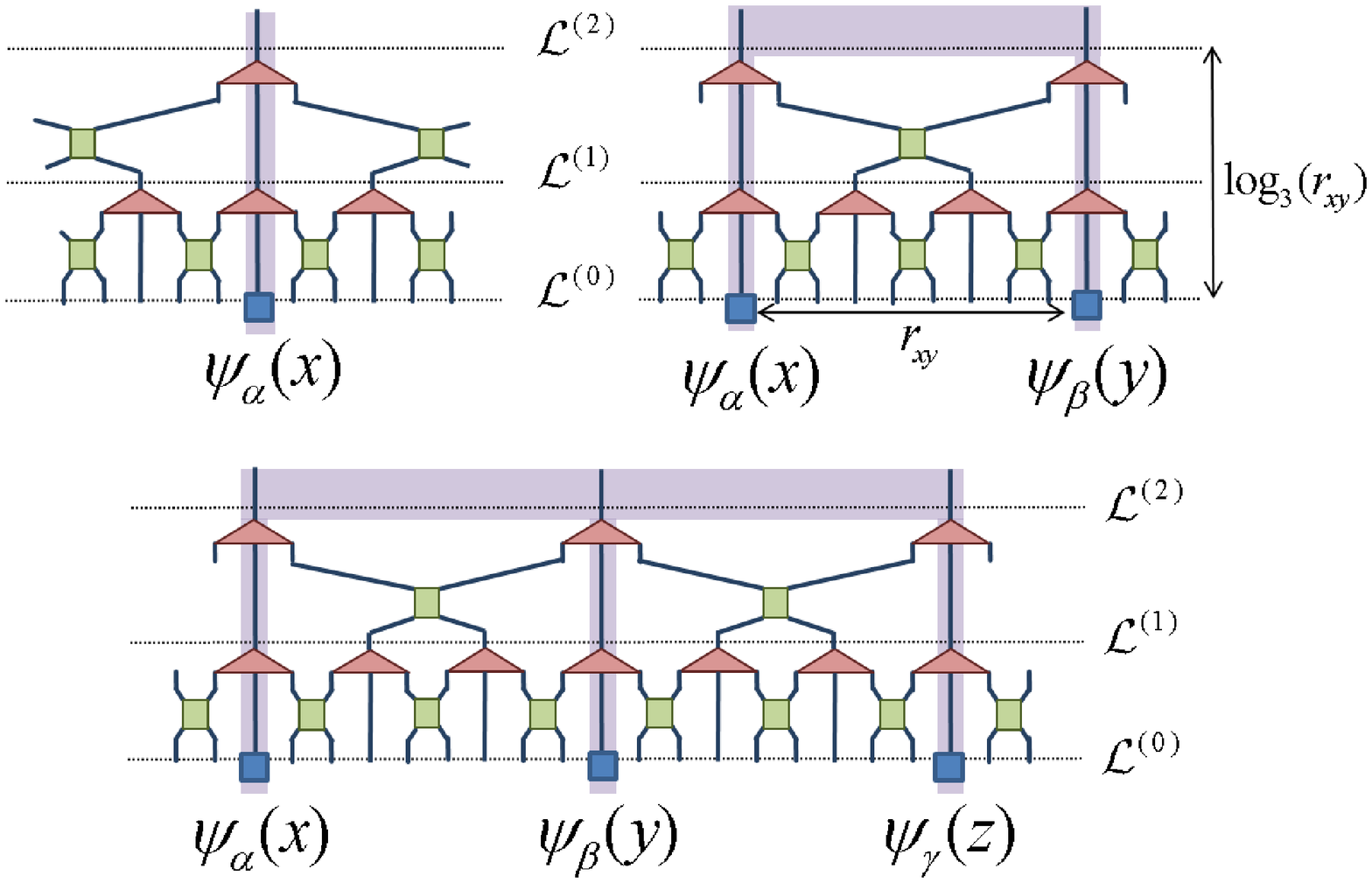}
\end{center}
\caption{One-point, two-point and three-point correlators for one-site scaling operators. After $\log_3 (r_{xy})$ iterations of the coarse-graining transformation, two operators separated by $r_{xy}$ sites become nearest neighbors. Hence the polynomial decay of correlations.
}
\label{V:fig:correlators}
\end{figure}
%%%%%%%%%%%%%%%%%%%%%%%%%%%%%%%%%%%%%%%%%%%

\subsection{Correlators and the operator product expansion}

Extracting correlators from a scale invariant MERA is also relatively easy. In terms of the one-site scaling operators $\psi_{\alpha}$ (see \cite{V:Pfeifer09} for details), one-point, two-point and three-point correlators read, 
\begin{eqnarray}
	&&~~~~~~\langle \psi_{\alpha} (x)\rangle = C_{\alpha}, ~~~~~~~~~~~~~	\langle \psi_{\alpha}(x) \psi_{\beta}(y) \rangle = \frac{C_{\alpha\beta}}{{r_{xy}}^{\Delta_{\alpha}+\Delta_{\beta}}}, \label{V:eq:2point}\\
	&&\langle \psi_{\alpha}(x) \psi_{\beta}(y) \psi_{\gamma}(z) \rangle = \frac{C_{\alpha\beta\gamma}}{
{r_{xy}}^{\Delta_{\alpha}+\Delta_{\beta} -\Delta_{\gamma}}
{r_{yz}}^{\Delta_{\beta}+\Delta_{\gamma} -\Delta_{\alpha}}
{r_{zx}}^{\Delta_{\gamma}+\Delta_{\alpha} -\Delta_{\beta}}
},~~~~\label{V:eq:3point}
\end{eqnarray}where the sites $x,y,z$ have been chosen conveniently and $r_{xy} \equiv |x-y|$, $r_{yz} \equiv |y-z|$ and $r_{xz} \equiv |x-z|$. The coefficients $C_{\alpha}$, $C_{\alpha\beta}$ and $C_{\alpha\beta\gamma}$ are given by
\begin{eqnarray}
&& C_{\alpha} \equiv \tr(\psi_{\alpha}\hat{\rho}) = \delta_{\alpha\mathbb{I}}, ~~~~~~~~~~
C_{\alpha\beta} \equiv \tr\left( (\psi_{\alpha}\otimes\psi_{\beta}) \hat{\rho} \right),\label{V:eq:Cab}\\
&& ~~~~~C_{\alpha\beta\gamma} \equiv 2^{\Delta_{\alpha}+\Delta_{\gamma}-\Delta_{\beta}} \tr\left( (\psi_{\alpha}\otimes\psi_{\beta}\otimes\psi_{\gamma}) \hat{\rho}\right),
\label{V:eq:Cabc}
\end{eqnarray}where $\hat{\rho}$ simultaneously denotes the reduced density matrix on one, two and three sites respectively. Thus, as expected in a critical system, the scale invariant MERA produces polynomial correlators. How this occurs is very intuitive. Consider for instance the two-point correlator $\langle \psi_{\alpha}(x) \psi_{\beta}(y) \rangle$ for $r_{xy} = 3^{\tau}$. As illustrated in Fig. \ref{V:fig:correlators}, operators $\psi_{\alpha}$ and $\psi_{\beta}$ become nearest neighbors after $\tau = \log_3 (r_{xy})$ iterations of the coarse-graining transformation. Since this transformation maps $\psi_{\alpha}$ into $\psi_{\alpha}' = 3^{-\Delta_{\alpha}} \psi_{\alpha}$, each iteration contributes a factor $3^{-\Delta_{\alpha}-\Delta_{\beta}}$ to the correlator, with
\begin{equation}
	(3^{- \Delta_{\alpha} - \Delta_{\beta}})^{\tau} = (3^{- \Delta_{\alpha} - \Delta_{\beta}})^{\log_3 (r_{xy})} 
	= 3^{\log_3({r_{xy}}^{-\Delta_{\alpha}-\Delta_{\beta}})} 
= \frac{1}{{r_{xy}}^{\Delta_{\alpha}+\Delta_{\beta}}},
\end{equation}
which explains its scaling. Finally, in order for the correlator $\langle \psi_{\alpha}(x) \psi_{\beta}(y) \rangle$ to be non-zero, the tensor product of the scaling operators $\psi_{\alpha}\otimes\psi_{\beta}$ must fuse into the two-site identity operator $\mathbb{I}$, which occurs with amplitude $C_{\alpha\beta}$.

The coefficients of Eqs. \ref{V:eq:Cab}-\ref{V:eq:Cabc} are analogous to those that appear in CFT. A proper choice of normalization of the fields leads to $C_{\alpha\beta}=\delta_{\alpha\beta}$, whereas coefficients $C_{\alpha\beta\gamma}$ define the operator product expansion (OPE), which for the primary fields of the Ising CFT reads
\begin{eqnarray}
	C^{\mbox{\tiny CFT}}_{\alpha\beta \mathbb{I}}\!= \!\delta_{\alpha\beta}, ~
  C^{\mbox{\tiny CFT}}_{\sigma\sigma\epsilon} \! = \frac{1}{2}, ~~
	C^{\mbox{\tiny CFT}}_{\sigma\sigma\sigma}\! = \!C^{\mbox{\tiny CFT}}_{\epsilon\epsilon\epsilon}\! =\! C^{\mbox{\tiny CFT}}_{\epsilon\epsilon\sigma}\! = 0.
	\label{V:eq:OPE_Ising}
\end{eqnarray}
A MERA with $\chi=16$ reproduces all these OPE coefficients with errors smaller than $3\times 10^{-4}$. 

The above comparisons with exact results from CFT allows us to conclude that entanglement renormalization produces accurate estimates of the critical properties of a system. Once this has been established, the present approach can be used to actually identify which CFT describes a given quantum critical point by estimating the conformal data (central charge, conformal dimensions and OPE of primary fields) that completely characterize it \cite{V:Pfeifer09}. The present discussion also indicates that the scale invariant MERA can be regarded as approximately realizing an infinite dimensional representation of the Virasoro algebra \cite{V:Henkel99}. The finite value of $\chi$ implies that only a finite number of the quasi-primary fields of the theory can be included in the description. Fields with small scaling dimension, such as primary fields, are retained foremost. However, an exact description requires $\chi$ to be infinite.

\subsection{Boundary critical phenomena}

\begin{figure}
\begin{center}
\includegraphics[width=12cm]{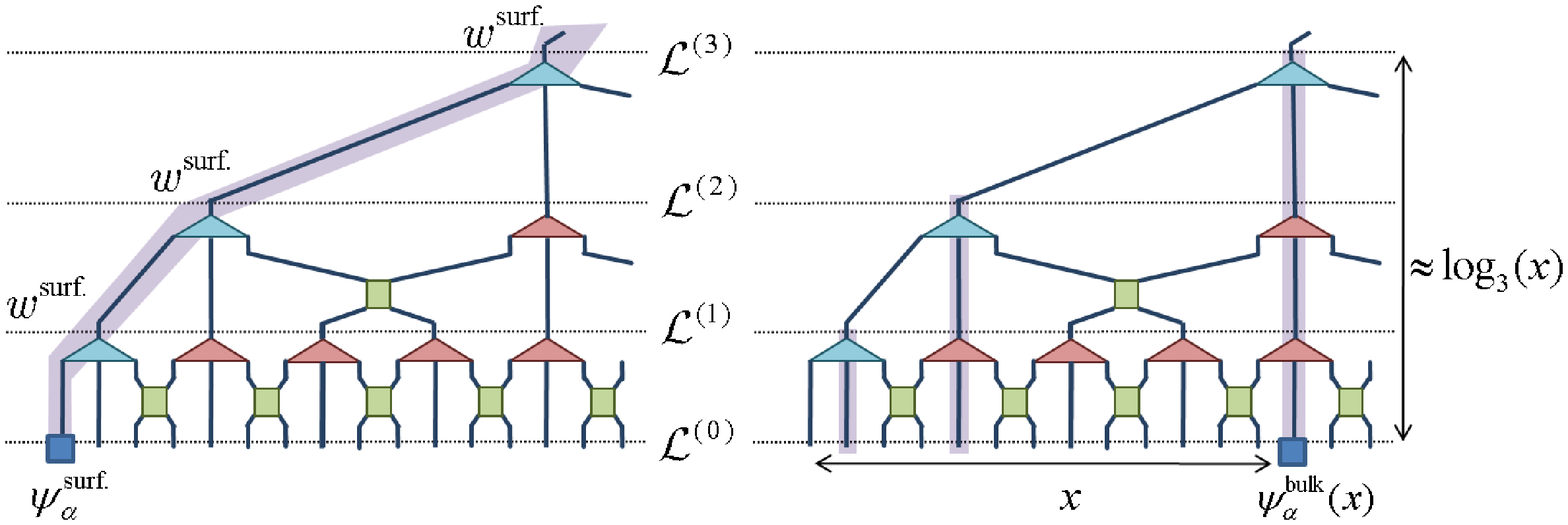}
\end{center}
\caption{Scale invariant MERA with a boundary made of copies of the same boundary isometry $w^{\mbox{\tiny surf.}}$ and the tensors $(u,w)$ of the bulk. 
}
\label{V:fig:boundary}
\end{figure}

So far we have restricted our attention to translation invariant systems. Critical systems with a boundary can also be described by adding a boundary to the scale invariant MERA \cite{V:Evenbly09d}. This is done by introducing a boundary isometry $w^{\mbox{\tiny surf.}}$ at the boundary, while the pair $(u,w)$ corresponding to the translation invariant case still represents the bulk, see Fig. \ref{V:fig:boundary}.

%%%%%%%%%%%%%%%%%%%%%%%%%%%%%%%%%%%%%%%%%%%
%%%%%%%%%%%%%%%%%%%%%%%%%%%%%%%%%%%%%%%%%%%

\begin{table} [!h]
\caption{Scaling dimensions for free and fixed boundary conditions}
\begin{tabular}{||l|l|l||l|l|l||}
  \hline
  $~~~~~\Delta_{\mbox{\tiny free}}^{\mbox{\tiny BCFT}}$  & $~\Delta^{\mbox{\tiny MERA}}_{\mbox{\tiny $\chi=16$}}~$ & error &

  $~~~\Delta_{\mbox{\tiny fixed}}^{\mbox{\tiny BCFT}}$ & $~\Delta^{\mbox{\tiny MERA}}_{\mbox{\tiny $\chi=16$}}~$ & error\\ \hline

   $(\mathbb{I})~$0 &  $~$0     & -- & $(\mathbb{I})~$0  &  $~$ 0      & --\\

   $(\sigma)$ 0.5 &  $~$0.499  & 0.2$\%$  & $~~~\,$ 2  &  $~$1.992   & 0.4$\%$\\

   $~~~\,$ 1.5 $~$  &  $~$1.503  & 0.18$\%$ & $~~~\,$ 3  &  $~$2.998   & 0.07$\%$\\

   $~~~\,$ 2   $~$  &  $~$2.001  & 0.07$\%$ & $~~~\,$ 4  &  $~$4.005   & 0.12$\%$\\

   $~~~\,$ 2.5 $~$  &  $~$2.553   & 2.1$\%$ & $~~~\,$ 4  &  $~$4.062   & 1.5$\%$\\
  \hline
  \end{tabular} \nonumber
  \label{V:table:scaling2}
 \end{table}

The presence of a boundary modifies the bulk correlators of Eqs. \ref{V:eq:2point}-\ref{V:eq:3point}. For instance, the expected value $\langle \psi_{\alpha} (x)\rangle$ becomes non-trivial
\begin{equation}
		\langle \psi_{\alpha} (x)\rangle_{\mbox{\tiny surf.}} \approx \frac{C^{\mbox{\tiny surf.}}_{0 \alpha}}{x^{\Delta_{\alpha}}}, \label{V:eq:1pointsurf}\\
\end{equation}
where $x$ is the distance to the boundary (in number of sites) and $C^{\mbox{\tiny surf.}}_{0 \alpha}$ is some constant. Again, this result is very intuitive. Consider $x = (3^{\tau+1}-1)/2$. As illustrated in Fig. \ref{V:fig:boundary}, after $\tau = \log_3 (\frac{2x+1}{3}) \approx \log_3 (x)$ iterations of the coarse-graining transformation the bulk scaling operator $\psi^{\mbox{\tiny bulk}}_{\alpha}$ has reached the boundary, into which it fuses with amplitude $C^{\mbox{\tiny surf.}}_{0 \alpha}$. In addition, by diagonalizing the \emph{boundary scaling superoperator} $\mathcal{S}^{\mbox{\tiny surf.}}$ that maps the boundary into itself at different length scales, we can extract the boundary scaling operators $\psi^{\mbox{\tiny surf.}}_{\alpha}$ and their scaling dimensions, as well as the fusion rules between bulk and boundary operators. Fig. \ref{V:fig:ScaleDim} and Table \ref{V:table:scaling2} show the boundary scaling dimensions for \emph{free} and \emph{fixed} boundary conditions. One can see that the number of conformal towers is smaller than in the bulk: for free boundary conditions, the primary fields left are the identity $\mathbb{I}$ and the spin $\sigma$, whereas for fixed boundary conditions only the identity $\mathbb{I}$ remains. Once more, the results compare well with the exact solution provided by (boundary) CFT \cite{V:Henkel99}.

\section{Summary \& Outlook}
\label{V:sect:Outlook}

In this chapter we have reviewed, step by step, the construction of a coarse-graining transformation for quantum many-body systems on a lattice that fulfills two natural requirements: it disposes of all short-distance degrees of freedom while it retains the properties of the ground state of a local Hamiltonian. This is achieved by using disentanglers $u$, that remove short-range entanglement across the boundary of a block of sites, and isometries $w$, that coarse-grain these blocks of sites according to White's rule and thus preserve the support of the ground state reduced density matrix.

This transformation can be used to compute ground state expected values of the form $\langle o_1 o_2 \cdots o_k \rangle$ both in finite and infinite systems. In a finite system, we simply coarse-grain the lattice until the effective Hamiltonian can be numerically diagonalized. In an infinite system, we iterate the coarse-graining transformation until we become sufficiently close to a fixed point of the RG flow --- that is, until we have eliminated all irrelevant perturbations in the original Hamiltonian. Then we expand local operators in terms of the scaling operators of the theory, whose expected values can also be determined. In a symmetry-breaking phase, the fixed point corresponds to an unentangled ground state, and local operators can be trivially evaluated. In a critical phase, the fixed point corresponds to a highly entangled ground state, but a scale invariant MERA can still be used to approximately compute the scaling operators and evaluate their expected values.

Notice that the scale invariant MERA allows us to make precise the notion of scale invariance in a lattice system. In order to get rid of the dependence on the original lattice spacing, one typically invokes the continuum limit. Here, instead, we have considered a coarse-graining such that all short-distance degrees of freedom are consistently discarded. The ground state of the lattice is then invariant under changes of scale if its effective description is locally identical to the original one. We envisage that this lattice version of scale invariance will 
become a useful testing ground for ideas and problems that are harder to analyze in the continuum limit. Recent work on the holographic principle might be a first step in this direction \cite{V:Swingle09}.

%%%%%%%%%%%%%%%%%%%%%%%%%%%%%%%%%%%%%%%%%%%
\begin{figure}
\begin{center}
\includegraphics[width=8cm]{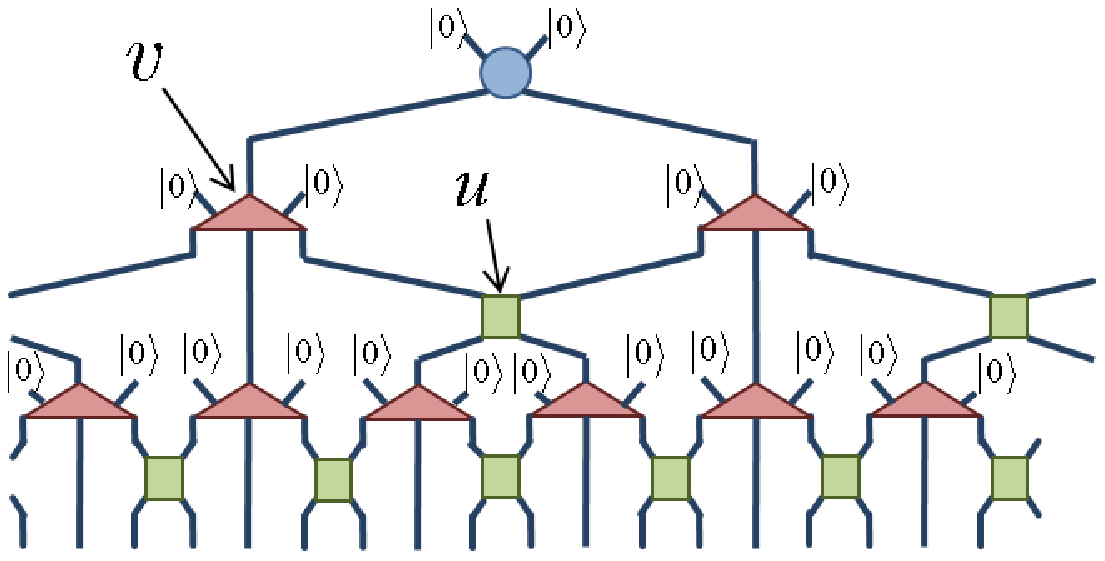}
\end{center}
\caption{The MERA seen as a unitary transformation $V$ such that $\ket{\GS} = V \ket{0}^{\otimes N}$.
}
\label{V:fig:QC}
\end{figure}
%%%%%%%%%%%%%%%%%%%%%%%%%%%%%%%%%%%%%%%%%%%

We conclude this chapter with a suggestive interpretation of the MERA. Notice that an isometry $w: \mathbb{V} \rightarrow \mathbb{V}^{\otimes 3}$ can be regarded as a unitary transformation $v: \mathbb{V}^{\otimes 3} \rightarrow \mathbb{V}^{\otimes 3}$ on three sites where two of these sites are initially in some fixed state $\ket{0}$, that is $w = v(\ket{0}\otimes\ket{0})$. The MERA can thus also be understood as a quantum circuit, made of gates $v$ and $u$, that implements a unitary transformation $V$ such that an unentangled state $\ket{0}^{\otimes N}$ becomes the ground state $\ket{\GS}$ of a local Hamiltonian $H$, see Fig. \ref{V:fig:QC}.  Thus, ground states of local Hamiltonians can be regarded as the result of a quantum computation that flows from the largest available length scales (i.e., the size of the system) to the smallest length scales (distance between sites of $\mathcal{L}$). At each length scale, unentangled wires in state $\ket{0}$ are added and subsequently entangled with the rest of the system by means of the gates $v$ and $u$.

The quantum circuit interpretation is quite insightful. For instance, it provides an explicit recipe to experimentally prepare the state $\ket{\GS}$ by entangling quantum systems, and it has been instrumental in generalizing the MERA to deal with fermionic and anyonic degrees of freedom \cite{V:Corboz09, V:Corboz09b, V:Eisert09, V:Eisert09b, V:Aguado09}. It also highlights the \emph{reversible} character of the coarse-graining transformation. During coarse-graining, short-distance degrees of freedom are not actually `thrown away', but rather `stored' in disentanglers and isometries, and can be re-incorporated in the picture later on. Indeed, as we have seen, from the reduced density matrix describing the state of the system at a given length scale, we can recover the state at a smaller length scale by using the descending superoperator $\mathcal{D}$. Finally, this viewpoint emphasizes the astonishing \emph{structural similarities} shared by most ground states of local Hamiltonians. Consider two systems in two completely different symmetry-breaking phases but with the same correlation length $\xi$. Both ground states become essentially disentangled after $O(\log (\xi))$ iterations of the coarse-graining. Therefore, the two ground states can be prepared from the same unentangled state by using the same quantum circuit, made of $O(\log(\xi))$ rows of identically wired gates, just by modifying the specific gates $v$ and $u$ applied in each case. 

Acknowledgments: The author thanks Glen Evenbly, Rob Pfeifer and Luca Tagliacozzo for stimulating conversations that helped improve this paper. Supported by the Australian Research Council (FF0668731, DP0878830).


\begin{thebibliography}{100}
 
\bibitem{V:Kadanoff67} 
L.P. Kadanoff, Rev. Mod. Phys. 39, 395 (1967).
%Static Phenomena Near Critical Points: Theory and Experiment.

\bibitem{V:Wilson75} 
K.G. Wilson, Rev. Mod. Phys. 47, 773 (1975).
%The renormalization group: Critical phenomena and the Kondo problem.

\bibitem{V:Fisher98} 
M.E. Fisher, Rev. Mod. Phys. 70, 653 (1998).
%Renormalization group theory: Its basis and formulation in statistical physics.

\bibitem{V:Cardy02} \emph{Scaling and Renormalization in Statistical Physics}, J. Cardy (Cambridge University Press, 1996).

\bibitem{V:Vidal07} 
G. Vidal, Phys. Rev. Lett. 99, 220405 (2007).
% Entanglement renormalization.
 
\bibitem{V:Vidal08} 
G. Vidal, Phys. Rev. Lett. 101, 110501 (2008).
% A class of quantum many-body states that can be efficiently simulated. 
 
\bibitem{V:Dawson08}
C.M. Dawson, J. Eisert, and T.J. Osborne, 
Phys. Rev. Lett. 100, 130501 (2008).
% Unifying variational methods for simulating quantum many-body systems  

\bibitem{V:Rizzi08}
M. Rizzi, S. Montangero, and G. Vidal, Phys. Rev. A 77, 052328 (2008).
%Simulation of time evolution with the MERA 

\bibitem{V:Evenbly09} 
G. Evenbly and G. Vidal, Phys. Rev. B 79, 144108 (2009).
%Algorithms for entanglement renormalization.
   

\bibitem{V:Evenbly07}
G. Evenbly and G. Vidal, arXiv:0710.0692.
%Entanglement renormalization in free fermionic systems 
 
\bibitem{V:Evenbly08}
G. Evenbly and G. Vidal, arXiv:0801.2449.
%Entanglement renormalization in free bosonic systems 

\bibitem{V:Cincio08} 
L. Cincio, J. Dziarmaga, and M.M. Rams, Phys. Rev. Lett. 100, 240603 (2008). 
%Multi-scale Entanglement Renormalization Ansatz in Two Dimensions: Quantum 

\bibitem{V:Evenbly09b} 
G. Evenbly and G. Vidal, Phys. Rev. Lett. 102, 180406 (2009).
%Entanglement renormalization in two spatial dimensions.

\bibitem{V:Evenbly09c} 
G. Evenbly and G. Vidal, arXiv:0904.3383. 
%Frustrated antiferromagnets with entanglement renormalization: ground state of the spin-1/2 Heisenberg model on a kagome lattice.

\bibitem{V:Aguado08} 
M. Aguado, G. Vidal, Phys. Rev. Lett. 100, 070404 (2008).
%Entanglement renormalization and topological order.  

\bibitem{V:Koenig08} 
R. Koenig, B.W. Reichardt, and G. Vidal, Phys. Rev. B 79, 195123 (2009).
%Exact entanglement renormalization for string-net models.
  
\bibitem{V:Giovannetti08} 
V. Giovannetti, S. Montangero, and R. Fazio, Phys. Rev. Lett. 101, 180503 (2008).
%Quantum MERA Channels. 

\bibitem{V:Pfeifer09} 
R.N.C. Pfeifer, G. Evenbly, and G. Vidal, Phys. Rev. A 79(4), 040301(R) (2009).   
% Entanglement renormalization, scale invariance, and quantum criticality.

\bibitem{V:Montangero08} 
S. Montangero, M. Rizzi, V. Giovannetti, and R. Fazio, Phys. Rev. B 80, 113103 (2009).
%arXiv:0810.1414.
%Critical exponents of one-dimensional quantum critical models by means of MERA tensor network.
 
\bibitem{V:Giovannetti09} 
V. Giovannetti, S. Montangero, M. Rizzi, and R. Fazio, Phys. Rev. A \textbf{79}, 052314 (2009).
%arXiv:0901.4424.  
%Homogeneous MERA states: an information theoretical analysis.  

\bibitem{V:Evenbly09d} 
G. Evenbly, R. N. C. Pfeifer, V. Pico, S. Iblisdir, L. Tagliacozzo, I. P. McCulloch, and G. Vidal, arXiv:0912.1642 
% surface critical phenomena with entanglement renormalization

\bibitem{V:Corboz09} 
P. Corboz, G. Evenbly, F. Verstraete, and G.Vidal, arXiv:0904.4151. 
%Simulation of interacting fermions with entanglement renormalization.

\bibitem{V:Eisert09} C. Pineda, T. Barthel, and J. Eisert, arXiv:0905.0669.

\bibitem{V:Corboz09b}
P. Corboz, and G.Vidal, Phys. Rev. B 80, 165129 (2009).

\bibitem{V:Eisert09b}T. Barthel, C. Pineda, and J. Eisert, Phys. Rev. A 80, 042333 (2009).
 
\bibitem{V:Aguado09} 
M. Aguado et al., \emph{in preparation}.
% anyonic MERA 

\bibitem{V:White92} 
S.R. White, Phys. Rev. Lett. {\bf 69}, 2863 (1992).
%; Phys. Rev. B {\bf 48}, 10345 (1993). 
%Density matrix formulation of quantum renormalization groups.

\bibitem{V:Tagliacozzo09}
L. Tagliacozzo and G. Vidal, arXiv:0903.5017.
%Simulation of two-dimensional quantum systems using a tree tensor network: entropic area law at work. 

\bibitem{V:Henkel99} 
\emph{Conformal Invariance and Critical Phenomena}, M. Henkel, (Springer, 1999).

%\bibitem{V:Zamolodchikov86} A.B. Zamolodchikov, JETP Lett. 43 (1986) 730.

\bibitem{V:Swingle09}
B. Swingle, arXiv:0905.1317.
%Entanglement Renormalization and Holography.
 

\end{thebibliography}
\end{document}